\documentclass[sigconf,screen,authorversion]{acmart}
\usepackage[utf8]{inputenc}
\usepackage{makecell}
 \usepackage{acmart-taps}
\usepackage{subcaption}
\AtBeginDocument{%
  }

\copyrightyear{}
\acmYear{2025}
\setcopyright{none}
\setcctype{}
\acmConference[CHI '25]{CHI Conference on Human Factors in Computing Systems}{April 26-May 1, 2025}{Yokohama, Japan}
\acmBooktitle{CHI Conference on Human Factors in Computing Systems (CHI '25), April 26-May 1, 2025, Yokohama, Japan}\acmDOI{10.1145/3706598.3713241}
\begin{document}

\title[How Users Provide Feedback to Shape Personalized Recommendation Content]{Beyond Explicit and Implicit: How Users Provide Feedback to Shape Personalized Recommendation Content}%Beyond Explicit and Implicit: How Users Leverage Feedback Mechanisms to Influence Personalized Recommendations

\author{Wenqi Li}
\affiliation{%
  \institution{Department of Information Management, Peking University}
  \city{Beijing}
  \country{China}
}\email{wenqili@pku.edu.cn}

\author{Jui-Ching Kuo}
\affiliation{%
  \institution{National Tsing Hua University}
  \city{Hsinchu}
  \country{Taiwan}}
\email{rosieeee321@gmail.com}

\author{Manyu Sheng}
\affiliation{%
  \institution{University of Chinese Academy of Sciences}
  \city{Beijing}
  \country{China}
}\email{shengmanyu@mail.las.ac.cn}

\author{Pengyi Zhang}
\affiliation{%
  \institution{Department of Information Management, Peking University}
  \city{Beijing}
  \country{China}
}\email{pengyi@pku.edu.cn}

\author{Qunfang Wu}
\affiliation{%
  \institution{Harvard University}
  \city{Cambridge}
  \state{MA}
  \country{USA}
}
\email{qunfangwu@fas.harvard.edu}
\renewcommand{\shortauthors}{Li et al.}

\begin{abstract}
As personalized recommendation algorithms become integral to social media platforms, users are increasingly aware of their ability to influence recommendation content. However, limited research has explored how users provide feedback through their behaviors and platform mechanisms to shape the recommendation content. We conducted semi-structured interviews with 34 active users of algorithmic-driven social media platforms (e.g., Xiaohongshu, Douyin). In addition to explicit and implicit feedback, this study introduced \textit{intentional implicit feedback}, highlighting the actions users intentionally took to refine recommendation content through perceived feedback mechanisms. Additionally, choices of feedback behaviors were found to align with specific purposes. Explicit feedback was primarily used for feed customization, while unintentional implicit feedback was more linked to content consumption. Intentional implicit feedback was employed for multiple purposes, particularly in increasing content diversity and improving recommendation relevance. This work underscores the user intention dimension in the explicit-implicit feedback dichotomy and offers insights for designing personalized recommendation feedback that better responds to users' needs.
\end{abstract}





\sloppy
\maketitle

% \section{Terminology Communication}
% We want to use terminologies consistently throughout our writing. For literature reviews, we will keep the terminologies as they were originally used in the literature.

% Below are several terminology options used in related papers:

% “Recommendation algorithms for social media feeds” in the paper “TikTok and the Art of Personalization: Investigating Exploration and Exploitation on Social Media Feeds.”

% “Recommendation systems” and “Recommendation algorithm” in the paper “Recommending to Strategic Users.”

% “Recommendation algorithms” and “personalized recommendation” in the paper “Measuring Strategization in Recommendation: Users Adapt Their Behavior to Shape Future Content.”

\section{Introduction}
% 需要添加explicit/implicit feedback 的解释，这是从系统角度的。Abstract里边也需要有
% 第二段要顺承故事引入feedback，第三段要介绍feedback定义。
% 以前的研究包括folk theory，imaginary等和我们是相似的，但我们的不同是在什么地方

Ariela is a design undergraduate in China and an avid social media user. She is captivated by the personalized feeds on Xiaohongshu\footnote{Xiaohongshu, also known as RedNote in English, is a community-sharing application. It can be accessed at \url{https://www.xiaohongshu.com}.}, which are filled with fashion trends, travel tips, and skincare products that guide her purchases and help her explore new hobbies (see \autoref{fig:feedback} \& \autoref{fig:xhspost} for Xiaohongshu's user interfaces). With each interaction, the platform seems to understand her better, continually delivering content that aligns with her preferences. Meanwhile, Ariela uses Douyin, the Chinese version of TikTok, for light entertainment during her breaks (see \autoref{fig:douyinmain} for Douyin's user interface). She enjoys watching short, fun videos on Douyin. When the algorithm suggests content she dislikes, she swipes past it, subtly guiding the platform toward content like cats or funny skits. However, Ariela soon realizes her feeds on both platforms, once diverse, are becoming repetitive, reflecting only her past choices. To break free from these patterns, Ariela deliberately searches for new styles and trends on Xiaohongshu or clicks ``Not interested'' on certain posts, disrupting its usual suggestions. And on Douyin, she fast skips videos she typically enjoys, hoping to inform the platform to offer something different.
\aptLtoX[graphic=no,type=html]{\begin{figure*}
    \centering
    \begin{subfigure}[t]{0.3\textwidth}
        \centering
        \includegraphics[width=0.6\textwidth]{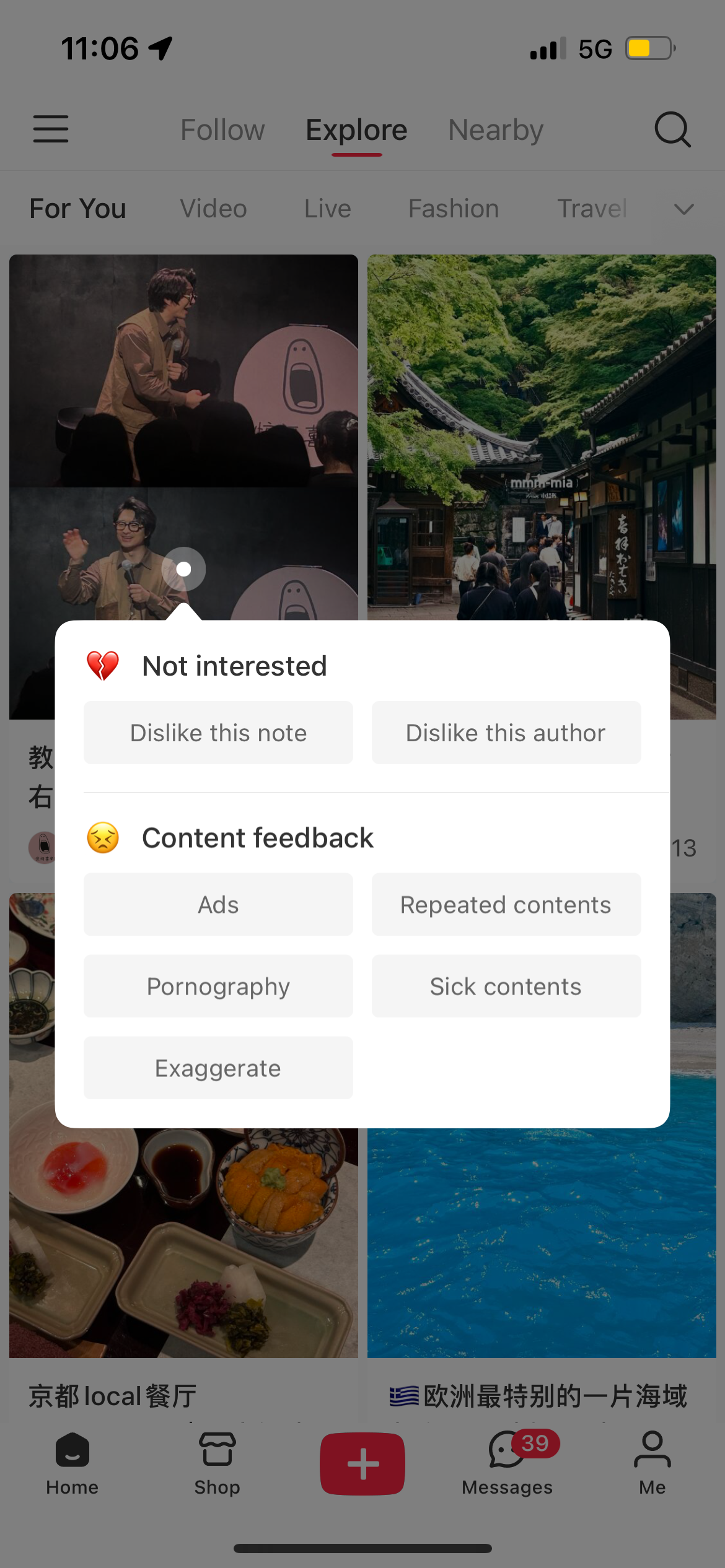}
        \caption{The main page of Xiaohongshu ``Explore'' and feedback options}
        \label{fig:feedback}
    \end{subfigure}
    \hfill
    \begin{subfigure}[t]{0.3\textwidth}
        \centering
        \includegraphics[width=0.6\textwidth]{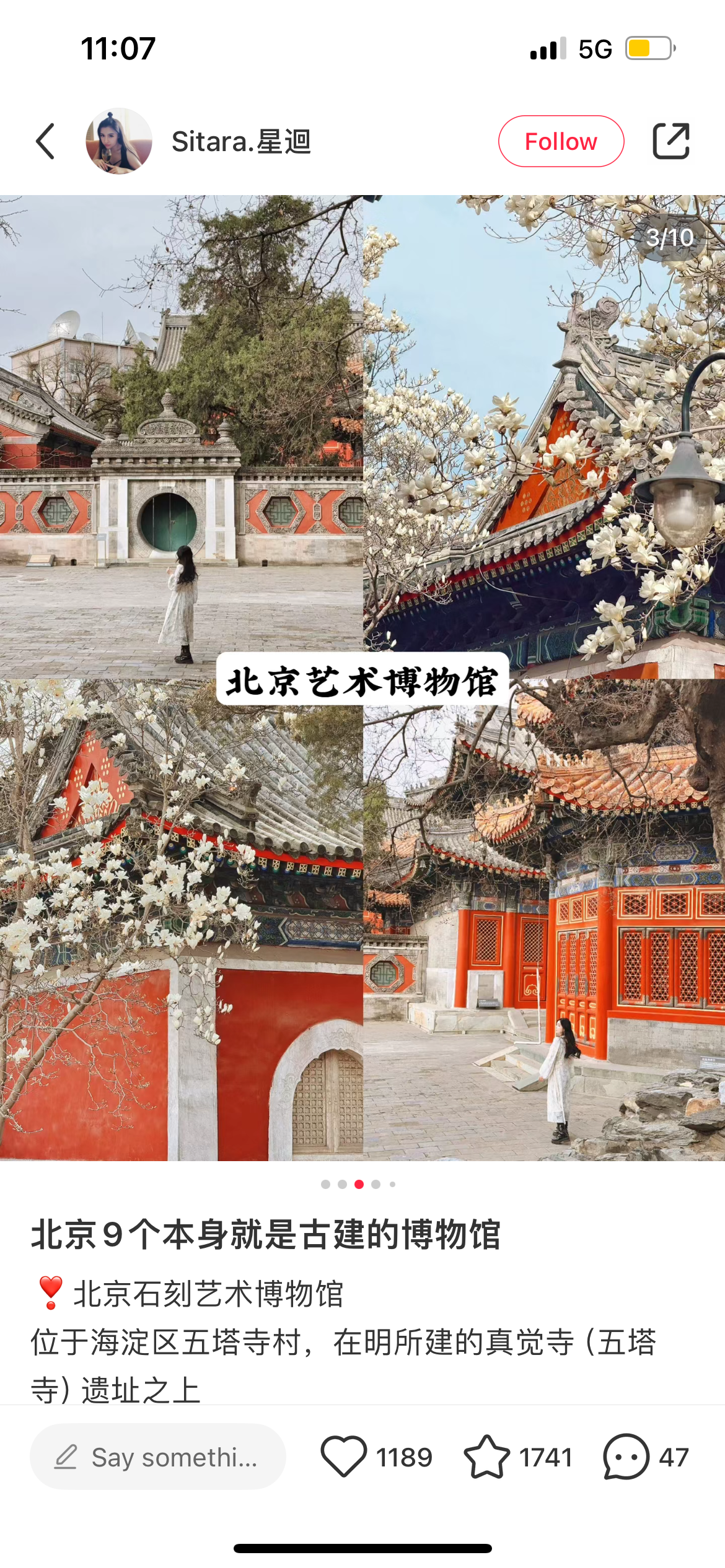}
        \caption{A Xiaohongshu post}
        \label{fig:xhspost}
    \end{subfigure}
    \hfill
    \begin{subfigure}[t]{0.3\textwidth}
        \centering
        \includegraphics[width=0.6\textwidth]{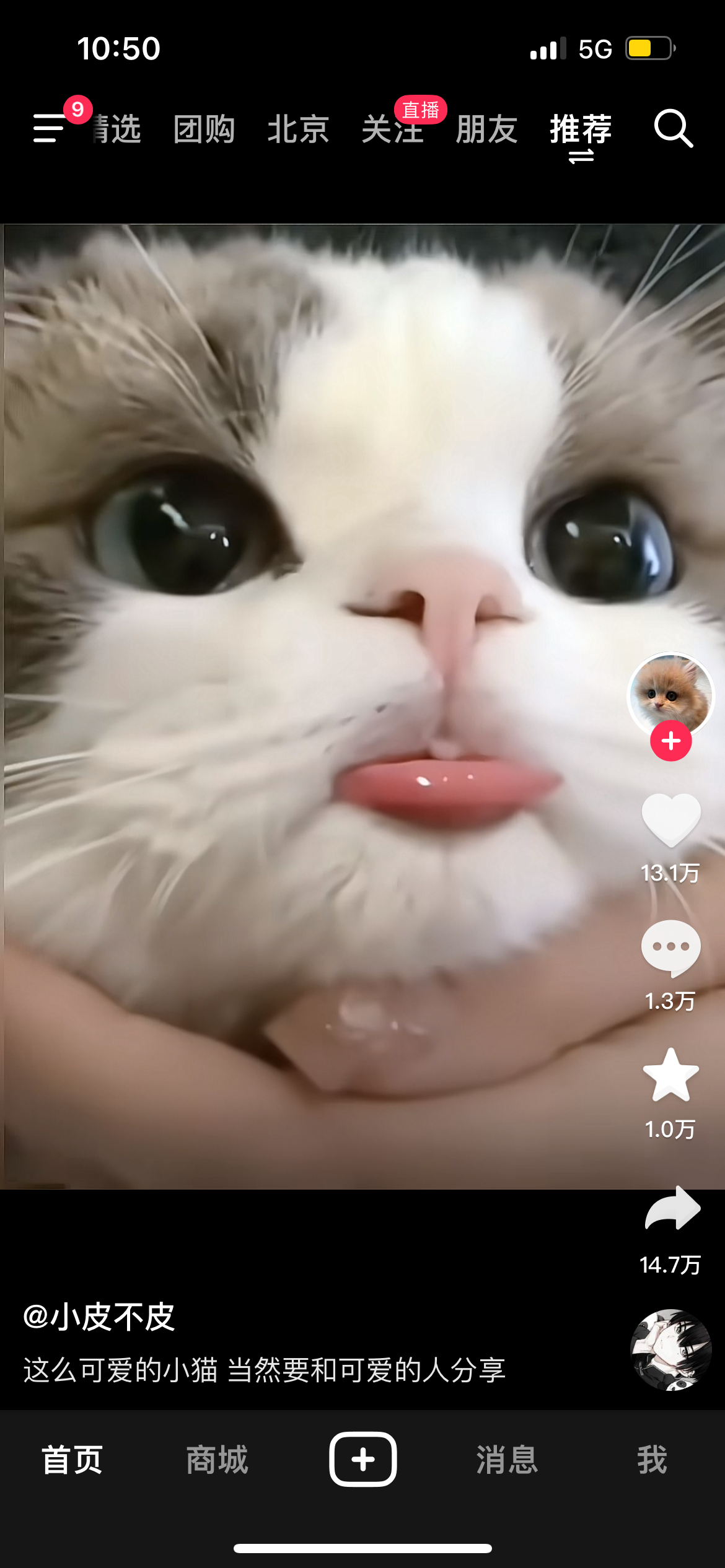}
        \caption{The main page of Douyin: ``For You''}
        \label{fig:douyinmain}
    \end{subfigure}
    \caption{The main user interfaces of Xiaohongshu and Douyin. 
    (a) is the main page of Xiaohongshu, the ``Explore'' page, displaying a selection of posts recommended by the algorithm, consisting of both picture and video posts. Located at the top of the main page is the search bar. Below the search bar, users often see trending hashtags and their interested channels that can be customized. Long press on a post can trigger options for reporting a post, including ``Not interested'' and ``Content feedback.''  
    (b) is a note detail page, where the note itself is the centerpiece. It typically includes a mix of text, images (or videos), and hashtags. Users can follow the creator, like, collect, or leave comments on the post. 
    (c) is the main page of Douyin, the ``For You'' page, showcasing a continuous stream of short videos curated by Douyin's algorithm for each user. A new video automatically displays as users scroll (or swipe) vertically. Users can like a video, leave a comment, share it, or follow the creator using the buttons on the right side. Douyin also provides similar content feedback features.}
    \label{fig:interface}
\vspace*{-10pt}
\end{figure*}}{\begin{figure*}
    \centering
    \begin{subfigure}[t]{0.3\textwidth}
        \centering
        \includegraphics[width=0.6\textwidth]{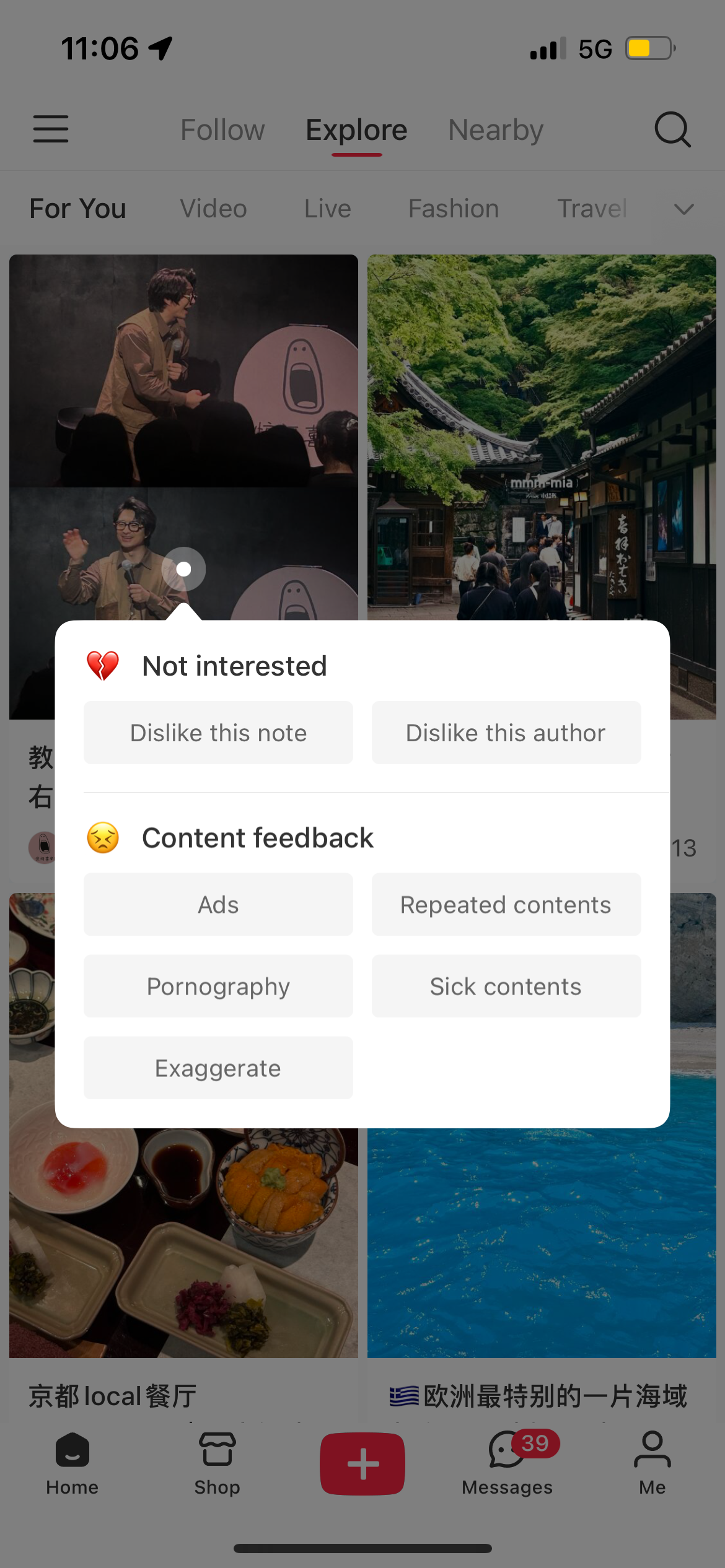}
        \caption{The main page of Xiaohongshu ``Explore'' and feedback options}
        \label{fig:feedback}
    \end{subfigure}
    \hfill
    \begin{subfigure}[t]{0.3\textwidth}
        \centering
        \includegraphics[width=0.6\textwidth]{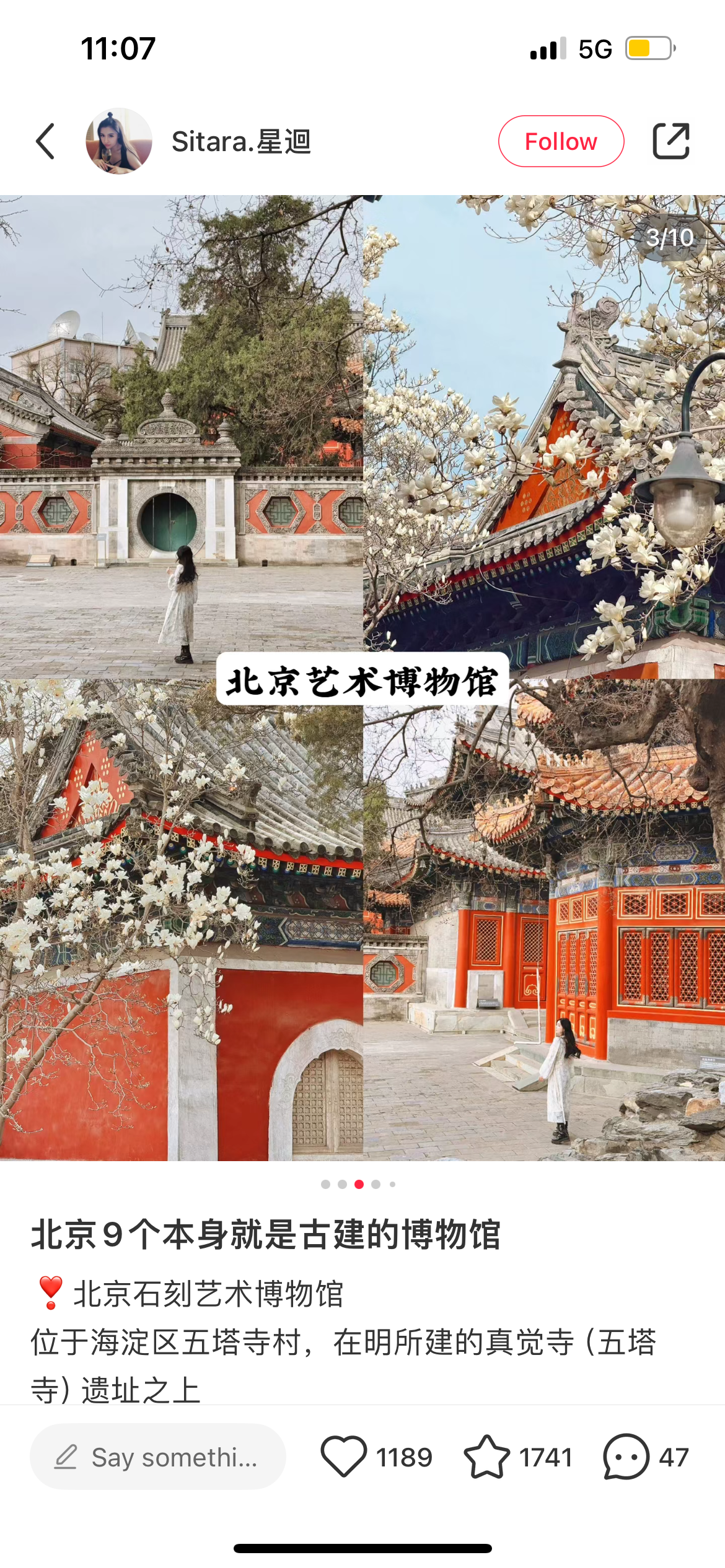}
        \caption{A Xiaohongshu post}
        \label{fig:xhspost}
    \end{subfigure}
    \hfill
    \begin{subfigure}[t]{0.3\textwidth}
        \centering
        \includegraphics[width=0.6\textwidth]{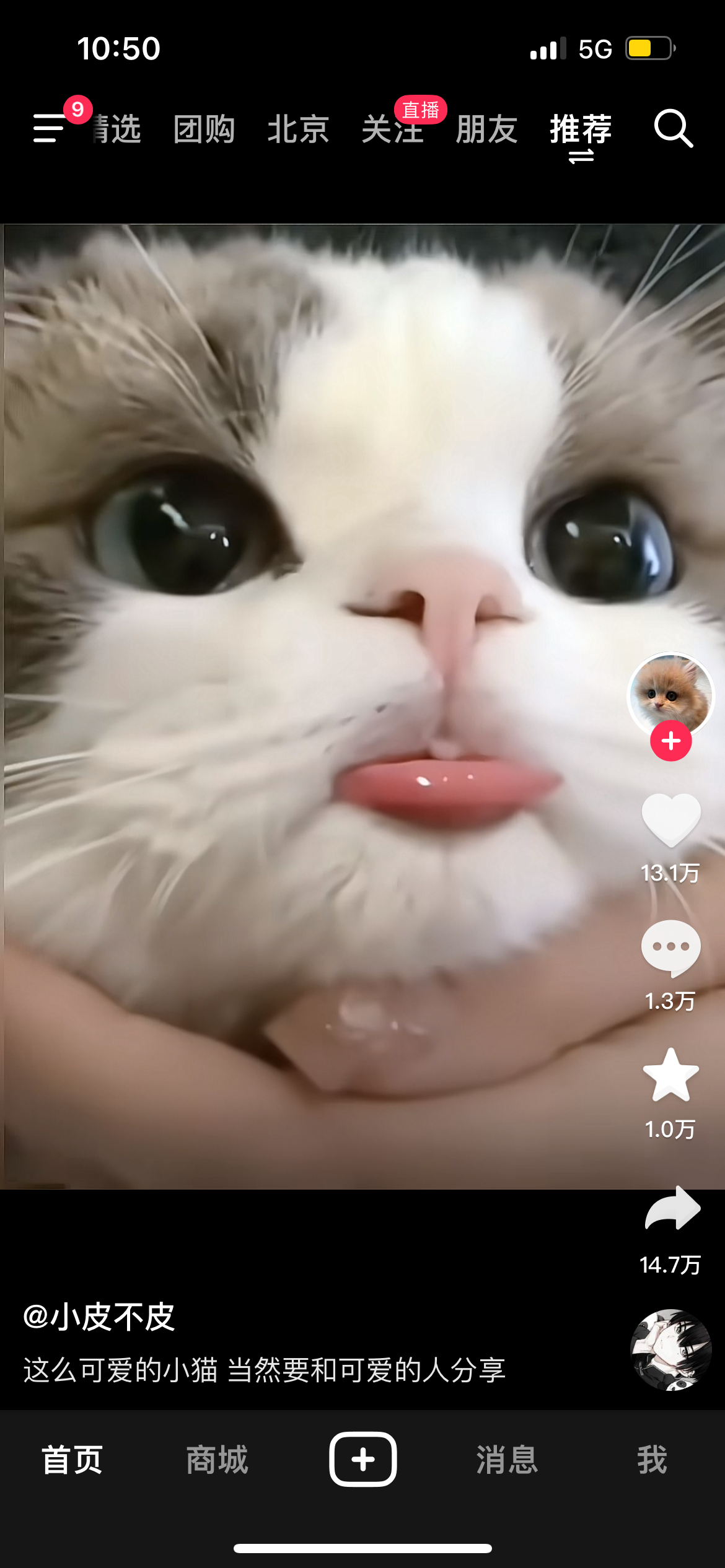}
        \caption{The main page of Douyin: ``For You''}
        \label{fig:douyinmain}
    \end{subfigure}
    \caption{The main user interfaces of Xiaohongshu and Douyin. 
    (\subref{fig:feedback}) is the main page of Xiaohongshu, the ``Explore'' page, displaying a selection of posts recommended by the algorithm, consisting of both picture and video posts. Located at the top of the main page is the search bar. Below the search bar, users often see trending hashtags and their interested channels that can be customized. Long press on a post can trigger options for reporting a post, including ``Not interested'' and ``Content feedback.''  
    (\subref{fig:xhspost}) is a note detail page, where the note itself is the centerpiece. It typically includes a mix of text, images (or videos), and hashtags. Users can follow the creator, like, collect, or leave comments on the post. 
    (\subref{fig:douyinmain}) is the main page of Douyin, the ``For You'' page, showcasing a continuous stream of short videos curated by Douyin's algorithm for each user. A new video automatically displays as you scroll (or swipe) vertically. Users can like a video, leave a comment, share it, or follow the creator using the buttons on the right side. Douyin also provides similar content feedback features.}
    \label{fig:interface}
\vspace*{-10pt}
\end{figure*}}

Just like Ariela, we increasingly consume content curated by personalized recommendation algorithms on social media platforms ~\cite{guy2010social}, such as Xiaohongshu, Douyin, and more. These algorithms, designed to capture user preferences through every click, view, or interaction, create a profile for each user to recommend content that is not only relevant but also engaging~\cite{setyani2019exploring}, enticing users into enduring usage ~\cite{seaver2019captivating}. Users form understandings of these algorithms through folk theories---``intuitive, informal theories that individuals develop to explain the outcomes, effects, or consequences of technological systems''~\cite{devito2017algorithms}---which affect how they interact with algorithms ~\cite{devito2017algorithms, devito2018people, eslami2016first, ngo2022exploring}. As personalized recommendation algorithms increasingly penetrate users' online and offline activities, concerns have emerged about the platforms ``spying'' on their preferences ~\cite{ellison2020we, klug2021trick} or pushing them into the homogeneous ``echo chamber'' ~\cite{gao2023echo} or ``information cocoons'' ~\cite{li2022exploratory}. In response, users have developed various strategies to influence the content recommended to them, such as refraining from hitting likes, tapping ``Not interested,'' searching for certain topics, or ignoring content they liked~\cite{kim2023investigating, Cen_Ilyas_Allen_Li_Madry_2024}. These strategies are based on the assumption that users’ behaviors will be captured by platforms as \textit{feedback} to the algorithms and, consequently, influence future recommendations~\cite{Cen_Ilyas_Allen_Li_Madry_2024}.

In the context of system engineering, feedback has been extensively studied to enhance the performance of information retrieval and recommender systems~\cite{kelly2003implicit}. These systems rely on both explicit and implicit feedback, a well-established dichotomy in the existing literature~\cite{jawaheer2014modeling}. Briefly, \textit{explicit feedback} refers to direct input provided by users to express their preferences, such as specifying keywords, rating, or answering questions about their interests, whereas \textit{implicit feedback} refers to various user interactions with the system, such as viewing, selecting, saving, or forwarding content, from which the system indirectly infers user preferences ~\cite{kelly2003implicit,jannach2018recommending}. As both concepts require user behavior or interaction as input, we argue that on personalized recommendation platforms, users strategically employing their behaviors to shape their recommendation feeds constitutes a form of feedback to the systems.

%Yet, due to lack of transparency in these algorithms ~\cite{gillespie2014relevance, pasquale2011restoring}, 
% 
In fact, user's strategization in algorithmic systems has been explored in the HCI communities. For example, scholars have also investigated user resistance to algorithms~\cite{karizat2021algorithmic, rosenblat2016algorithmic} and user autonomy~\cite{ngo2022exploring, kang2022ai, feng2024mapping}. While these studies highlight users' intentions and potential to influence recommendation feeds, limited research has connected users' perceptions with system feedback mechanisms. Studying the connection could lead to improved feedback designs in personalized recommendations. To address this gap, our work began with a broader inquiry: \textit{How do users provide feedback through their behaviors and platform mechanisms to shape and control the content presented to them on personalized recommendation platforms?}
% However, as personalized recommendation becoming more sophisticated, they also raise important concerns about user autonomy and the transparency of algorithmic processes ~\cite{ngo2022exploring, kang2022ai,feng2024mapping}.

We conducted semi-structured interviews with 34 active users of personalized recommendation platforms (e.g., Xiaohongshu and Douyin). We found that users employed a variety of feedback mechanisms to influence the content they receive. These mechanisms range from explicit feedback, such as marking content as ``Not interested,'' to implicit feedback like clicking and liking, from which user preferences were indirectly inferred. We found that the traditional explicit-implicit dichotomy~\cite{kelly2003implicit, jannach2018recommending} failed to fully capture users' agency when they consciously employ behaviors previously categorized as implicit feedback to shape their recommendation feeds. To address this, we divided the implicit feedback category to \textit{intentional implicit feedback} and \textit{unintentional implicit feedback}. Unlike the conventional understanding of implicit feedback as passive or natural interactions, intentional implicit feedback refers to behaviors consciously performed by users with the expectation that the system will interpret them as signals of their preferences. For example, behaviors like quickly skipping a disliked post or deliberately clicking interested posts to get more related content were frequently observed in our study. These behaviors, categorized as implicit feedback in previous research, are distinctly intentional in nature. At the same time, unlike \textit{explicit feedback}, intentional implicit feedback allows users to guide their recommendations without direct input to express their preferences. By introducing the \textit{intention} dimension into the dichotomy, we highlight users' proactive engagement in shaping their feeds through both explicit and implicit feedback behaviors. By recognizing intentional implicit feedback, platforms can more accurately capture user intent and provide users with a greater sense of control over their feeds. 

Additionally, we found that users' feedback behaviors were closely associated with their purposes. Explicit feedback was primarily used for feed customization goals like reducing inappropriate content and improving recommendation relevance, while intentional implicit feedback emerged as crucial for feed customization to increase content diversity and improve recommendation relevance. Unintentional implicit feedback was most commonly linked to content consumption. The findings underscore the need to better design for implicit feedback in personalized recommendations and to align feedback mechanisms with users' specific purposes.

This work has the following contributions: first, the study provided empirical evidence of how users leveraged different feedback mechanisms to fulfill their purposes when using personalized recommendation platforms. Second, the work introduced the concept of intentional implicit feedback, expanding beyond the traditional explicit and implicit feedback dichotomy. The intentional implicit feedback captures user's intentions in taking their actions to influence algorithms and therefore future recommendation feeds. Third, the study offers design implications for personalized recommendation platforms to support more transparent and purpose-oriented feedback mechanisms. 
\section{Related Work}
\subsection{User Perceptions of Personalized Recommendations}
% What is personalized recommendation platforms;
% Xiaohongshu and TikTok
% How existing literature studied users in personalized recommendation? For example, folk theories studies; user strategization.
% What is the gap/challenges left in existing literature?

Personalized recommendation platforms utilize algorithms to tailor content to individual users based on their preferences and behaviors, such as subscriptions, clicks, likes and dislikes, and dwell time ~\cite{adomavicius2005toward, setyani2019exploring,yi2014beyond}. Previously, recommender systems have been widely used in search engines, news consumption, and e-commerce sites \cite{schafer1999recommender,lu2015recommender,raza2022news}. Powered by recommendation algorithms, recommender systems capture and analyze user interactions, such as clicks, purchases, or explicit ratings, to build user models that represent preferences and behaviors of users. Based on these models, recommendation algorithms like collaborative filtering analyze similarities between users or items to generate personalized recommendations that match each user's profile~\cite{zanker2009case}. More recently, social media platforms are increasingly integrating personalized recommendation algorithms~\cite{guy2010social} to keep users engaged for longer periods of time by delivering content aligned closely with their preferences, as well as facilitating users’ content creation and networking ~\cite{seaver2019captivating,van2018networks, kang2022ai}. As a result, personalized recommendation algorithms are gradually taking up editorial roles in shaping what users view and know ~\cite{gillespie2014relevance}. They have reshaped the content consumption ~\cite{bucher2016algorithmic}, content creation ~\cite{devito2018people, bucher2018cleavage}, and online networking ~\cite{eriksson2021algorithmic} in social media.
Platforms like Douyin and Xiaohongshu are gaining tremendous popularity among domestic Chinese users as well as international audiences (known as TikTok and RedNote). These kinds of platforms allow users to directly interact with the content and rely heavily on algorithms to capture these interactions, rather than solely on account follows, to optimize individually customized image or video feeds ~\cite{Huang_2021, klug2021trick, Chen2019}. 

However, due to the opaque nature of the underlying algorithms ~\cite{gillespie2014relevance, pasquale2011restoring}, users have very limited understanding of how the personalized recommendation platforms operate ~\cite{eslami2015always}. This lack of understanding often leads users to develop ``folk theories'' ~\cite{devito2017algorithms, eslami2016first, ngo2022exploring, bucher2016algorithmic} to make sense of how the systems function to tailor contents delivered to them. Klug et al. found that TikTok users, for instance, often assume that video engagement, posting time, and the accumulation of hashtags are key factors that influence the platform's algorithm recommendation~\cite{klug2021trick}. Such folk theories are not static; instead, they evolve as users encounter new experiences and information, which helps them navigate their interactions with algorithmic systems ~\cite{devito2018people}. These folk theories directly influence how users perceive and interact with the algorithms, based on which users would exert control over the algorithm by taking action to improve content personalization ~\cite{haupt2023recommending} or increase their visibility on social media platforms ~\cite{burrell2019users, bucher2018cleavage}. Content creators also share and discuss their experiences with algorithms, namely ``algorithmic gossip,'' to collectively refine their strategies in promoting content consumption~\cite{bishop2019managing}. 

To sum, personalized recommendation platforms have reshaped how users consume, create, and share content online. But, users often lack a clear understanding of how their interactions are processed by the platform. As users form ``folk theories'' to understand the underlying mechanism, it becomes crucial to explore how users interact with these personalized recommendation platforms and provide feedback to shape their recommendation feeds, which leads to the next section of the literature review.

\subsection{User Strategic Interaction with Recommendation Algorithms}
Users are becoming more aware that their interaction behaviors could influence algorithms and further shape their online experiences, sometimes resulting an uncomfortable feeling that personalized recommendation algorithms are ``spying'' on their thoughts~\cite{ellison2020we,klug2021trick}. Such awareness lead to a variety of user behaviors aimed at teaching, resisting, and repurposing algorithms ~\cite{kim2023investigating} as well as personalizing content moderation ~\cite{jhaver2023personalizing}. 

Some of these behaviors are subtle content modification actions, such as ``voldemorting'' (i.e., not mentioning words or names) and ``screenshotting'' (i.e., making content visible without sending its website traffic) to control their online presence~\cite{van2018networks} or ``algospeak'' (i.e., intentionally altering or substituting words when creating or sharing online content) to bypass algorithmic moderation ~\cite{klug2023how}. Additionally, users spend sustained time and effort to combat unwanted recommendation content using various strategies ~\cite{papadamou2022just, ricks2022does}, such as refraining from hitting likes, tapping ``Not interested,'' searching new keywords~\cite{kim2023investigating}, ignoring content they liked~\cite{Cen_Ilyas_Allen_Li_Madry_2024}, and configuring personalized content moderation tools by blocking specific keywords~\cite{jhaver2022designing}. Some research challenged algorithmic recommendation and content moderation systems that perpetuate inequalities and injustices ~\cite{karizat2021algorithmic,bishop2018anxiety,rosenblat2016algorithmic,kasy2021fairness}. For example, TikTok users modified their engagement---such as following users and sharing their contents---to influence their recommendation feeds, align them with their personal identities, and also impact other users’ feeds to resist the suppression of marginalized social identities ~\cite{karizat2021algorithmic}. They also engaged with specific hashtags and likes to curate a desired ``For You'' feed in response to perceived inequalities ~\cite{simpson2021}. To avoid incorrect moderation in social media, users used coded language or stopped using the platforms when they perceived that the platform disproportionately removed
marginalized users’ identity-related content ~\cite{mayworm2024content}, and sometimes they resorted to switching account after being shadowbanned~\cite{yao2024blind}. User-driven algorithm auditing is also leveraged to uncover harmful algorithmic behaviors ~\cite{devos2022toward, shen2021everyday}. 

Thus, algorithms are now shaped not only through users’ organic interactions with the platform, but also through their strategic attempts to influence the recommendation feeds ~\cite{lee2022algorithmic}. For example, Haupt et al. modeled this strategic process as a two-stage noisy signaling game, where users first strategically consume content presented to them in an initial ``cold start'' phase to influence future recommendation feeds, and then, the system refines its suggestions based on these interactions, leading to an equilibrium where user preferences are clearly distinguished ~\cite{haupt2023recommending}. Taylor and Choi extended the human-algorithm interaction by adding that users notice the personalization and perceive the algorithm as responsive to their identity, which further shapes their interactions and outcomes on the platform ~\cite{taylor2022initial}. Some research also refer to users' purposeful manipulation as ``gaming'' the algorithm ~\cite{petre2019gaming,SHEPHERD2020102572, hardt2016strategic}. Content creators may ``play the visibility game'' by leveraging relational and simulated influence to gain commercial benefits~\cite{cotter2019playing}. While gaming could spur innovations and discover new uses for existing platforms ~\cite{bambauer2018algorithm}, the strategic adaptation of behaviors may as well be misinterpreted by the algorithm and degrade its accuracy ~\cite{Cen_Ilyas_Allen_Li_Madry_2024}.

These studies indicate that users consciously employ a variety of behaviors to influence algorithms and shape their recommendation feeds, guided by their perceptions or ``folk theories'' of how the algorithms work. However, limited research has connected these user perceptions with the platforms' underlying feedback mechanisms. Understanding how users' perceptions align with or diverge from the algorithm's intended responses---and how they interact to shape recommendation feeds---can inform improvements in feedback design for personalized recommendation systems.

\subsection{Implicit and Explicit User Feedback}
Feedback mechanisms play an important role in human-computer interaction. System feedback communicates the response to a user's action \cite{perez1996collaborative} , while user feedback can be used as a useful input for system optimization ~\cite{Spink/Losee:96}.  User interactions are captured and used as user feedback to refine recommendation feeds. In fields of information retrieval ~\cite{Spink/Losee:96} and recommender systems ~\cite{oard2001modeling}, these mechanisms are commonly divided into two categories: explicit and implicit feedback. \textit{Explicit feedback} relies on direct user input to specify their preferences, such as specifying keywords, marking documents, rating, or answering questions about their interests ~\cite{kelly2003implicit}. It usually represents a deliberate, unambiguous, and intentional quality assessment by a user ~\cite{jannach2018recommending}. However, explicit feedback mechanisms often impose a higher user cost, demanding additional effort beyond typical search behavior ~\cite{kelly2003implicit, gadanho2007addressing}. In contrast, \textit{implicit feedback} refers to all kinds of user interactions with the systems from which the system can indirectly infer user preferences ~\cite{jannach2018recommending}. It unobtrusively obtain user preferences based on their natural interactions, such as viewing, selecting, saving, and forwarding ~\cite{kelly2003implicit,jannach2018recommending}. 

Observable user behaviors used as feedback were categorized into several types, including examination, retention, social and public action, physical action, creation, and annotation, with explicit feedback behaviors mainly falling under the annotation category ~\cite{jannach2018recommending}. Among these behaviors, clicks have long been used as a major feedback mechanism by search engines as indicators of intent and relevance. Scholars attempted to analyze users' click behavior under search queries~\cite{zhang2011user,shen2012personalized} to optimize the ranking algorithms of search engines. The number of clicks on ads has also become the primary indicator of revenue for commercial search engines ~\cite{kim2011advertiser}. Because of the economic motivation, numerous studies have explored how to predict ad clicks more accurately in search engines ~\cite{jacques2015differentiation,liu2010personalized}. In social media platforms, more diverse traits are used to predict user preferences, such as the user's friend networks~\cite{ellison2020we}, posts~\cite{kim2016click}, likes~\cite{cheung2022influences, hu2022interest}, and bookmarks~\cite{klaisubun2007behavior}. Oard and Kim outlined the \textit{minimum scope} of different behaviors, which is the minimum level at which feedback is applied---whether it impacts a portion of an object (segment level, e.g., viewing comments of a post), an entire object (object level, e.g., collecting a post), or multiple objects (class level, e.g., platform-wide searches) ~\cite{oard2001modeling}. Jawaheer et al. compared different properties between implicit and explicit feedback. For instance, explicit feedback provides transparency but also required cognitive effort, thus only a small percentage of users actively provide it, often leading to bias toward those who are more expressive ~\cite{jawaheer2014modeling}. In terms of \textit{polarity}---the positive (preference) or negative (dislike) orientation of the feedback---explicit feedback contains both whereas implicit feedback is typically considered positive ~\cite{jawaheer2014modeling, hu2008collaborative}. These properties, along with factors like user goals ~\cite{nazari2022choice,liang2023enabling}, digital literacy ~\cite{devito2018people}, and task complexity ~\cite{white2005study}, all have potential influence on users' understanding and adoption of different feedback mechanisms. For example, skilled users rely on platform cues to grasp algorithms, while those with lower web skills depend on external guidance. Users with peacekeeping self-presentation goals tend to limit explicit hashtag use compared to those focused on authenticity ~\cite{devito2017algorithms}. Users are reluctant to provide explicit feedback in complex tasks but are more willing to do so in media and entertainment domains ~\cite{white2005study,jawaheer2014modeling}. Moreover, the interpretation and effectiveness of these feedback is also dependent on contextual factors including user characteristics ~\cite{jawaheer2014modeling} and system functionalities ~\cite{liu2024train}.

While the explicit-implicit dichotomy offers clarity in examining user feedback from the system’s perspective, it may not fully capture the complexity of users' intentions when interacting with personalized recommendation platforms. Users now consciously manipulate behaviors once considered natural and unobtrusive to influence recommendation feeds, blurring the line between explicit and implicit feedback. Particularly, the term ``implicit'' can refer to unintention, background attention, and lack of awareness, causing confusion about when an interaction is truly implicit versus explicit ~\cite{serim2019explicating}. 
This evolving landscape necessitates a combined perspective from both users and systems, as well as a re-examination of how and why users employ different feedback mechanisms in personalized recommendation systems.
\section{Methods}
To answer the research questions, we conducted a semi-structured interview study for 34 users who had used personalized recommendation platforms (i.e., Xiaohongshu, Douyin, Kuaishou\footnote{Kuaishou is a content and social platform known for its short video sharing and emphasis on authentic expression. It more focuses on community engagement and live-streaming interactions.}, and BiliBili Shorts\footnote{Bilibili is a Chinese video streaming platform where users can share and interact through video uploads and comments. Its mobile application offers a short video section similar to YouTube Shorts, where users can create and share brief, engaging videos, but with a stronger emphasis on anime and gaming due to its community nature.}). We used both inductive and deductive approaches to analyze the data. This study was approved by the Institutional Review Board of Syracuse University and conducted in compliance with ethical guidelines of the respective institutions of all authors.  

\subsection{Participant Recruitment}
We targeted active users who had used personalized recommendation platforms. We distributed pre-screening surveys through social media platforms, online forums, and personal networks. The eligibility criteria included (1) being 18 years or older and (2) having at least six months of experience using the personalized recommendation platforms. We asked participants to self-identify the personalized recommendation platforms they commonly used. The pre-screening surveys helped select a diverse group of participants, ensuring variation in gender, age, educational backgrounds, and daily usage patterns. Additionally, we employed snowball sampling, asking initial participants to refer other users who met the study's criteria. This approach was effective in reaching individuals with varying levels of engagement, from lurkers who primarily consumed content to heavy users who actively posted and interacted with content.

We recruited 34 participants in total. The majority of the participants were aged 18-25 ($n = 25$), with a higher representation of females ($n = 19$). Participants reported using multiple personalized recommendation platforms. Douyin ($n = 29$, 85\% of participants) and Xiaohongshu ($n = 24$, 71\% of participants) were the most commonly used; other platforms used included Bilibili Shorts ($n = 10$), Kuaishou ($n = 7$), and TikTok ($n = 1$). All these platforms provide personalized recommendation content including short videos, images, and texts. The primary interfaces of Xiaohongshu and Douyin are shown in \autoref{fig:interface}. Interfaces of Kuaishou (shown in \autoref{fig:ksmain} \& \autoref{fig:ksvideo} in \autoref{appendix:interface}) are similar to Xiaohongshu, and interfaces of Bilibili Shorts (shown in \autoref{fig:bili} in \autoref{appendix:interface}) are akin to Douyin. The duration of usage for their most frequently used platform ranged from less than a year to over four years. According to reports, the majority of users on these platforms are younger generation under 35~\cite{flow2023demographic,marketingtochina2024douyin,kuaishou2022user}, and over 70\% of Xiaohongshu's users are female ~\cite{flow2023demographic}. Our sample of participants exhibits an age and gender distribution that aligns with the general user demographics of these platforms. More participants information are presented in \autoref{tab:participants} in \autoref{appendix:demographics}.

\subsection{Interview Procedure}
We conducted semi-structured interviews from December 2022 to April 2023. Each interview lasted between 40 to 60 minutes. The interviews were conducted in person or via video conferencing platforms to accommodate participants' schedules and geographic locations. All interviews were conducted in Mandarin Chinese. Participants received a compensation of 25 Chinese Yuan. Participants were informed that they could withdraw from the study at any time without penalty.

The interview protocol was designed to explore participants' engagement with personalized recommendation platforms. First, we asked the participants about their platform usage, including the platforms they were using, the content they were interested in, and the interactions with content and platforms (e.g., browsing, posting, liking, and searching). Then, we asked their understanding of and attitudes toward personalized recommendation and platforms’ algorithms, as well as perceived impacts of platform use. Particularly, we delved into how participants responded to the algorithms, such as their strategies for managing content exposure and content preferences and avoiding undesirable content. We wrapped up the interviews by asking their overall suggestions to the design of personalized recommendation platforms.

Participants were informed about the study's goal and procedure before the interviews began. Each interview was audio-recorded with the participants' consent, and detailed notes were taken to capture key points. The recordings were transcribed verbatim and anonymized for analysis. 

\subsection{Data Analysis}
We conducted a codebook thematic analysis of the interview data~\cite{braun_thematic_2019,emeline_brule_thematic_2020}. We first conducted inductive coding, allowing themes to emerge from the data. We then incorporated deductive analysis by integrating feedback concepts from existing literature to further refine and apply the codes. Finally, we conducted code co-occurrence analysis~\cite{namey2008data} to explore the relationship between user feedback behaviors and purposes for interacting with platforms.

The data analysis was conducted locally on MAXQDA, with the codebook shared among researchers for discussion. All interviews were analyzed in Mandarin Chinese to retain the original nuances and meanings during interpretation. Selected quotes were then translated into English for presentation in this paper. 

In the inductive coding phase, our researchers first read through all transcripts to familiarize themselves with the data, and then each of them independently conducted an open coding for a different portion of the transcripts, allowing themes to emerge through constant comparison and memo-ing~\cite{miles1994qualitative}. Throughout this process, the researchers held regular discussions to compare their codes and refine the open coding scheme. They synthesized the codes into categories and composed an initial codebook. The categories include: user behaviors, interaction purposes, perceptions and attitudes, folk theories, and challenges of interacting with algorithms.

Transitioning to the deductive phase, we compared the coding results with existing literature. We identified that the ways users interacted with the algorithms could be interpreted into user providing feedback to recommendation systems. Based on existing literature, recommendation system feedback is usually categorized into explicit and implicit feedback~\cite{jannach2018recommending,jawaheer2014modeling,kelly2003implicit}. Explicit feedback requires users additional input beyond their normal behavior, such as rating and answering questions about their interests, while implicit feedback is derived unobtrusively from users' natural interactions with the system, such as viewing, selecting, and forwarding~\cite{kelly2003implicit}. We observed that, in our study, participants intentionally leveraged implicit feedback mechanisms (e.g., clicking a post) to provide feedback to their personalized recommendation. While some of the strategic behaviors to shape recommendation feeds have been documented in prior literature~\cite{Cen_Ilyas_Allen_Li_Madry_2024}, limited research has connected users' perceptions with system feedback mechanisms. 

By comparing to the literature, we refined the codebook. First, we narrowed down to focus on two code categories: \textit{interaction purposes} and \textit{user behaviors}. We referred to the established categories of \textit{explicit feedback} and \textit{implicit feedback} in ~\cite{jannach2018recommending} to categorize user interaction with systems. Then, we found that within implicit feedback behaviors, users consciously and proactively shape the recommendation feeds, which contradicts the original definition of implicit feedback. Therefore, we divided implicit feedback into \textit{intentional implicit feedback} and \textit{unintentional implicit feedback} to differentiate whether users' intention is present or absent during implicit feedback behaviors. Overall, we categorized users' feedback behaviors into three types:
\begin{itemize}
    \item \textit{Explicit feedback}: users' direct input to express their preferences or interests.
    
    \item \textit{Intentional implicit feedback}: behaviors that users consciously perform to influence recommendation content, with their knowledge that these actions might be interpreted by the platforms to infer their interests.
    
    \item \textit{Unintentional implicit feedback}: users' natural interactions with the platforms without any deliberate intention to influence recommendation content.
\end{itemize}

%Then, within each feedback categories, we refined the specific codes of user behaviors by referring to existing literature ~\cite{jannach2018recommending, Cen_Ilyas_Allen_Li_Madry_2024}, while retaining the codes of behaviors (e.g., initiate a new search, using or collecting hashtags, and ignoring or swiping past a post) that were identified in inductive coding and specific to our study context.

To further understand and interpret feedback behaviors, we mapped three key properties to each identified behavior: features (i.e., specific platform features such as the ``like'' button or search box that afford feedback behaviors), polarity ~\cite{jawaheer2014modeling,hu2008collaborative} (i.e., ``positive'' or ``negative'' feedback), and minimum scope ~\cite{oard2001modeling} (i.e., the minimum level at which feedback is applied: ``segment,'' ``object,'' or ``class''). We carefully analyzed all the mentioned platforms to identify the corresponding feature(s) for each behavior and assigned polarity and scope based on interpretation of participant transcripts and platform functionalities.

Then, two researchers completed the deductive coding based on the refined codebook. For the first eight transcripts, they coded them independently and reviewed the coding together to any discrepancies and refined the coding guidelines accordingly. This iterative process helped to ensure consistency in interpretations. Then, they independently coded the remaining transcripts, 13 transcripts for each. During the process, the two researchers continued to share summary memos and addressed any ambiguities during weekly discussions with the research team.

We identified potential correlation patterns between feedback behavior types and purposes for interacting with platforms (i.e., content consumption, directed information seeking, content creation and promotion, and feed customization) as well as specific feed customization sub-purposes. To explore their relationships, we used code co-occurrence analysis. Specifically, we identified instances where a user behavior code and a user purpose code appear together within the same interview segment. To avoid duplication, we counted each behavior-purpose co-occurrence only once per participant using MAXQDA. For example, if a behavior-purpose co-occurrence was mentioned multiple times in a single participant’s interview, it was treated as a single co-occurrence instance for that participant.  %By focusing on the co-occurrence of codes rather than individual cases, we ensured that our analysis reflected the behavior-purpose patterns across participants, rather than the frequency of mentions within one interview. 
We then aggregated these co-occurrence instances within each of the three feedback types, i.e., explicit feedback, intentional implicit feedback, and unintentional implicit feedback. 
We also conducted co-occurrence analysis between the sub-purposes (i.e., improving recommendation relevance, increasing content diversity, reducing inappropriate content, and protecting privacy) and user feedback behaviors. Notably, we only counted instances within intentional implicit feedback and explicit feedback for the sub-purposes, because participants did not specify unintentional implicit feedback behaviors corresponding to the four sub-purposes. 

\section{Findings} 
Our analyses revealed three types of user feedback and the ways this feedback aligns with users' purposes. We first present the three types of feedback: explicit feedback, intentional implicit feedback, and unintentional implicit feedback. 
%and how users leverage these feedback to achieve their purposes.
Then, we summarize four purposes for which users interacted with algorithms: content consumption, directed information seeking, content creation and promotion, and feed customization. Last, we showcase the relationships between purposes and the types of feedback adopted. We found that explicit feedback was primarily used for feed customization, while intentional implicit feedback emerged as crucial for feed customization, particularly in increasing content diversity and improving recommendation relevance. Unintentional implicit feedback was most commonly linked to content consumption and directed information seeking. We leverage the interview data to explain underlying motivations for adopting different feedback types to fulfill their purposes.

\subsection{Types of User Feedback for Personalized Recommendation Content}
% The primary difference between explicit and implicit feedback is whether users provide feedback to the system with or without their intentions. Our analyses suggests that within the implicit feedback category, users consciously leveraged implicit feedback mechanisms to customize the feeds. For example,. Thus, we considered a separate category, namely ``intentional implicit feedback,'' to highlight the active role users play in shaping their content feeds based on their knowledge of algorithms.

% \textit{Explicit feedback} refers to users' behaviors where they directly communicate their preferences and interests to personalized recommendation algorithms. \textit{Intentional implicit feedback} involves user behaviors that, while not explicitly stated as feedback, were consciously performed with the knowledge that these actions might be interpreted by the system to infer user interests. While the feedback is intentionally made by users, it remains implicit to the system, which continues to process them and infer user preferences as it would any implicit feedback. \textit{Unintentional implicit feedback} consists of user behaviors that allow algorithms to assess user preferences without any deliberate intention from the user to give feedback.

We identified six explicit feedback behaviors, nine intentional implicit feedback behaviors, and 13 unintentional implicit feedback behaviors. As mentioned in the Methods, intentional implicit feedback and unintentional implicit feedback were different in whether a user's behavior has the intention for providing feedback to the platforms or not. For example, users may like a post or follow a user out of appreciation, or they may do it to inform the algorithm their preferences. Given that unintentional implicit feedback is common and natural usage of platforms and had been studied as ``implicit feedback'' in a substantial body of prior research~\cite{kelly2003implicit,jannach2018recommending}, in this section, we focus on explaining explicit feedback and intentional implicit feedback. \autoref{tab:feedbackType} presents these behaviors along with corresponding features, polarity, acting scope, and the number of participants who reported them. \autoref{tab:unintentional} in \autoref{appendix:unintentional} presents the behaviors for unintentional implicit feedback, including creating, collecting, and sharing a post, commenting and viewing comments, browsing profile pages, purchasing, and more. %They were mostly common usage of the algorithm-mediated apps, and were already being studied by a substantial body of research, especially on implicit feedback.

\begin{table*}[]
\small
\caption{Summary of behaviors for different feedback types and their corresponding characteristics, including platform features involved, the polarity (positive or negative) of feedback, the scope of the behavior, and the number of participants.}
\label{tab:feedbackType}
\resizebox{\textwidth}{!}{%
\begin{tabular}{p{0.37\linewidth}p{0.18\linewidth}p{0.08\linewidth}p{0.14\linewidth}p{0.15\linewidth}}
\toprule
\textbf{Behavior} & \textbf{Features}& \textbf{Polarity} & \textbf{Minimum Scope} & \textbf{Participant Count} \\ \midrule
\textbf{Explicit Feedback} &  &  &  &  \\
Mark as not interested & Not interested & - & Object & 21 \\
Use/Follow hashtags & Collect hashtags* & + & Object & 4 \\
Report a post or user & Report & - & Object & 3 \\
Disable personalization & Personalization options & - & Class & 3 \\ 
Subscribe to interested topics & Choose your interests & + & Class & 3 \\
Block a user & Block & - & Object & 2 \\
\midrule
\textbf{Intentional Implicit Feedback} &  &  &  &  \\
Ignore or swipe past a post &  & - & Object & 20 \\
Initiate a new search & Search & + & Class & 19 \\
Click a post (or comments) to view &  & + & Object & 14 \\
Stop using the platform or switch platforms &  & - & Class & 12 \\
Dwell on a post &  & +  & Segment & 4 \\
Like a post & Like & + & Object & 2 \\
Follow a user & Follow & + & Object & 2 \\
Refresh the feed &  & - & Class & 2 \\
Talk for platform monitoring &  & + & Class & 1 \\
\bottomrule
\multicolumn{5}{l}{* denotes the behavior or the feature exclusive to the Xiaohongshu platform.}
\end{tabular}%
}
\vspace*{-10pt}
\end{table*}

\subsubsection{Explicit feedback}
The explicit feedback behaviors identified in our analyses occurred mostly at the object scope and were all supported by specific platform features. The most frequently used explicit feedback, reported by 21 participants, was marking a post as ``Not interested.'' Participants used it as negative feedback to inform the algorithm that they do not wish to see similar content in the future. P06 stated that this was \textit{``the most direct and easy way to express dislike, where you can proactively intervene with the algorithm with just one extra step.''} Participants provided various reasons for using the ``Not interested'' feature such as lack of interest in the post content, ads, and poor content quality. Sometimes expression of non-interest was more on granular and nuanced content level. For example, users might appreciate fashion content in general, but their tastes could vary widely, thus not all fashion-related posts were relevant to them (P08).
% 就有一段时间他我讲上一次我都不感兴趣，那是有段时间他老给我推送一个博主的穿搭视频，然后博主他穿搭是我完全不感兴趣的，他可能是那种成贵妇名媛风的那种感觉，然后我是不会喜我不是这种风格的
Several participants highlighted the importance of using this feature to reduce homogeneous content for a diverse and engaging feed. P13 noted that \textit{``more niche content might appear after marking many similar recommendations as `Not interested,' making your feed more diverse and encouraging you to spend more time on the platform.''}
% 但是有的时候你会发现，比如说你点了之后，他可能还是会举个例子，可能说我点了这个之后，你会发现他第二天还是出现，然后你可能第二天你还要点，你不是说你点一次他就能解决问题的，可能你要点好几次他才能解决问题。所以我理解可能是它这个系统可能也在试探你就试一下我减少一点，你有没有发现，然后再减少更多。所以其实我的理解你可能要一直的反馈，或者让他明确的感知到你其实不喜欢这个内容，然后他的那种才会逐渐的恢复到你喜欢的，但是你点一次肯定是没用，他还会再出现。
Although commonly used, some participants found that platforms did not respond to this feedback effectively or immediately. P12 mentioned that the effect was not satisfactory as \textit{``it stopped recommending closely related content, but continued to suggest somewhat related items,''} which led her to use it less frequently. Meanwhile, P08 and P12 observed that it often required multiple attempts before the algorithm significantly reduced similar content. P12 speculated that the platform required consistent feedback to make gradual adjustments:
\begin{quote}
    It's not a one-time solution... I think the platform is testing you by slightly reducing the frequency to see if you noticed, and then reducing it further. You need to keep providing feedback or make it clear that you really don't like this content, and then it will gradually adjust. (P12)
\end{quote}
% 但是有的时候你会发现，比如说你点了之后，他可能还是会举个例子，可能说我点了这个之后，你会发现他第二天还是出现，然后你可能第二天你还要点，你不是说你点一次他就能解决问题的，可能你要点好几次他才能解决问题。所以我理解可能是它这个系统可能也在试探你就试一下我减少一点，你有没有发现，然后再减少更多。所以其实我的理解你可能要一直的反馈，或者让他明确的感知到你其实不喜欢这个内容，然后他的那种才会逐渐的恢复到你喜欢的，但是你点一次肯定是没用，他还会再出现。
Other negative explicit feedback included blocking, reporting, and disabling personalization, which were less frequently used. Blocking was to prevent further interactions with specific users, effectively removing their content from the participant's feed. Reporting was to flag content to the platform's moderation team, signaling that the content violated community guidelines. Participants only resorted to blocking or reporting when they experienced strong negative reactions such as when the content was offensive and inappropriate. Unlike blocking and reporting that provide feedback to specific object (a post or user), disabling personalization operates at class scope to convey an overall dissatisfaction. Three participants had tried disabling the ``personalization options'' feature in Xiaohongshu, either due to privacy concerns or to avoid addiction by preventing the platform from over-learning their preferences. However, they eventually re-enabled it after finding the non-personalized feed much less relevant and engaging. As P01 reflected that his perspective changed after disabling and re-enabling personalization: 
\begin{quote}
    After disabling personalization, my Xiaohongshu usage dropped significantly as the content no longer appealed to me, so I turned the personalization back on. After re-enabling it, my perspective shifted. I realized that the benefit of having personalization on is that it saves a lot of time by providing targeted recommendations, especially when searching for something specific. (P01)
\end{quote}
% 其实就是第一是关闭个性化之后，也让我感受到确实在一定程度上小红书这个软件我的使用频率下降了，并且他推的一些文件不是文章或者说笔记都是我不喜欢的了，所以我又把个性化给打开了。 打开了之后和最开始开的过程又会有一个思路上的变化，就是会感觉其实打开个性化的好处就是它能帮你节省大量的时间，比如说你可能打开之后就是想搜这个东西，你就会发现小红书一直在非常垂直的给你推这个领域的东西，他会帮你去节省一些搜索类的时间，以及他能帮你快速的展露一些你想要看到的一些内容也好，或者一些其他的方面东西也好，反正一是他做的确实很垂直，第二是它的丰富度确实没有问题
While most explicit feedback was negative, participants mentioned two positive explicit feedback mechanisms. One was the use of hashtags to increase visibility when publishing posts, or to follow content of interest. On Xiaohongshu, users who were not content creators can collect specific hashtags to follow updates and receive recommendation content tagged with those hashtags. Yet P09 felt that the hashtag functions \textit{``mainly benefited content creators to drive traffic rather than the viewers.''} She wished the platform would make collecting hashtags at the class level more useful by providing notifications for new posts, especially in niche areas. 

The other positive explicit feedback was subscribing to topics of interest, which involved selecting preferred content categories or topics. This differs from following individual users or channels, as it informs the platform about users' general interests rather than subscribing to specific user-generated content. Though the platform allowed later edits, participants either did not notice this feature or felt it was unnecessary to make changes.

\subsubsection{Intentional implicit feedback} 
The majority of intentional implicit feedback behaviors were positive. These behaviors, without direct platform features to prompt users, mostly relied on users' spontaneous intention to provide feedback.

Ignoring (in Xiaohongshu) or swiping (in Douyin) past a post was the most common intentional implicit feedback observed by 20 participants. Ignoring a post was to deliberately not click certain posts in Xiaohongshu's recommendation feed. Participants would ``actively skip'' (P01, P07), ``filter out'' (P04), or ``not pay attention to'' (P11) posts they were not interested in. Participants often combined this behavior with its opposite---intentional click, a positive feedback behavior reported by 14 participants. Swiping past a post in Douyin was to fast skip a video without engaging with it. As a form of negative feedback to avoid uninterested or homogeneous content, several participants preferred ignoring or swiping past content over marking it as ``Not interested'' for its efficiency and subtlety. For instance, P25 considered some content was \textit{``facets of the real world''} that should not be marked, believing that even if the content was not personally engaging, it represented diverse aspects of society and reality. P07 believed that if many people marked content as ``Not interested,'' it might not reach those who needed it, so she chose to simply swipe past the content. Both P25 and P07 used ignoring or swiping past the content to convey \textit{subtle} negative feedback, as marking ``Not interested'' would cause the content to disappear immediately. Although this feedback was subtle, participants found it to be reasonably effective. For example, P27 said: \textit{``Swiping past is the quickest way I deal with videos I'm not interested in, and it (the platform) learns from that behavior.''}
% 如果说很多人都选择不感兴趣，他可能不会推给真正有需要的人，我觉得他这个帖子就会沉下去，但有的人是真的会需要这方面的信息，所以我不会选择不感兴趣，包括我也会看抖音抖音，他也会就是说让你选他也可以选不感兴趣，但是我就不太会学，我会直接视频就刷过去 (P07)

Another intentional implicit feedback frequently used by participants was to initiate a new search. This was a proactive approach to get more targeted recommendation content. Participants believed that searching for specific topics prompted the algorithm to update their profile and push more relevant content. Some participants also leveraged search, or clicking on irrelevant search recommendations, to seek diverse content or escape ``information cocoon'' (P22, P29, and P34). P12 stated that it could be viewed as negative feedback, indicating that the recommendation feeds were too constrained. She would \textit{``try to start a novel search to override the overwhelming content''} to inform the platform to show her something else.
% 然后我就有点下意识的想要搜索一些奇特的东西，我觉得能不能相当于是我在用我的行为告诉他，我不想看这也是想给我换别的这种感觉...我有的时候会企图用搜索来覆盖我之前就那些铺门盖地的东西 

While many participants observed obvious changes in recommendation feeds following search feedback, sometimes the algorithm's response might not be as accurate or immediate as intended. For example, P12 searched for exam centers but received various exam-related posts, such as good luck rituals, which she did not believe in but led to anxieties. P11 highlighted the algorithm's failure to capture her shift in interest:
\begin{quote}
   I was very into ``Haikyuu!!'' and searched for it many times. The frequent searches and repeated clicks on related content in the homepage led the algorithm to push a lot of ```Haikyuu!!'' mechandise. But then I moved on to another anime, the system continued to flood my feed with ``Haikyuu!!'' content. Even after I searched for the new anime, the platform still hadn't learned that my interest had shifted. (P11)
\end{quote}
% 包括后来我用小红书搜，我之前是很喜欢看排球少年，然后搜了排球少年，可能是由于次数过多，然后在首页也疯狂点击，就导致他开始给我推送非常多的关于排球少年的动漫周边之类的东西。
% 后来排球少年看完了，我又看了另外一部动漫，然后这时候他就完全干不过我首页的排球少年，就搜搜完之后，他还是没有出现在我的手艺，但其实我目前已经对排球少年没有那么大的热情了，这点就是小红书没有发现我的爱好已经转变了(P11)。 
Despite the frequent use of search to intentionally provide feedback, it was also naturally used for information seeking, which was listed as ``search for information'' in unintentional implicit feedback.

% Implicit search refers to search actions taken when users are intrigued by a particular post and seek further information. In particular, when users engage in directed information seeking, they continue to search for validation of the post. Several participants mentioned a feature provided by multiple platforms that prompts search queries for related content or presents a search link in the comment area suggesting trending searches. P18 explained:

% \begin{quote}
%     I often come across new online buzzwords or topics I'm unfamiliar with after watching a video. TikTok does a great job with this; under each video, if you open the comments section, there's a `Trending Searches' button at the top that you can click to quickly find explanations or more related content. This shows that it's not just me; many people search for related content after watching a video.
% \end{quote}
% Participant P11 reflected that following the search prompts will lead to more recommendations on related sub-genres.

There were 12 participants mentioning that they stopped using the platforms or switched to other platforms when they felt the recommendation content no longer aligned with their interests or became too repetitive. This disengagement could serve as negative feedback to \textit{``let the recommendation system cool down''} (P26). It also indicated a desire to regain control over their content consumption and avoid becoming overly dependent on algorithm-driven feeds. Some participants reported stop using or uninstalling apps when their feeds became monotonous. P19 observed that after watching a couple of OpenAI videos, the algorithm flooded him with AI related content. He managed this by temporarily stop using the app, and upon returning, the intensity of such recommendation contents had decreased. Others mentioned switching to platforms less reliant on personalized algorithms, seeking a fresh perspective or a break from repetitive content.

Other intentional implicit feedback included dwelling on or liking posts, following users, refreshing the feed, and even talking about certain topics to ``let the platform monitor.'' Participants intentionally adjusted their dwell time on a post to convey their preferences to platforms. For instance, watching a video to the end, or repeatedly, signaled strong interest. P08 noticed that the algorithm’s positive reinforcement from finishing a lengthy video was so strong that she had to mark related posts as ``Not interested'' multiple times to mitigate its influence. 
% 因为我可能就看了一次，我就点进去看了一次，然后他经常给我更新他的每条视频，我就很有点无语，然后我就给他点了不感兴趣，后来还给我推送了一次，然后我又点了一次不敢学，然后后来他就没给我退。 
% Researcher Z 11:19 
% 点了退了两次，它才消失在你的信息推送里面，你觉得他们为什么会假如说你说你才点了一次，然后他就老给你推，你觉得是为什么呢？ 
% 你之前有其他的也有这种情况吗？就只点了一次，然后就一直在推。 
% Participant 11:40 
% 有点记不住了，感觉好像没有，反正这是我印象比较深刻的一次，因为可能我当时看那个视频很长，但是我我给他看完了，因为我还看了评论，因为博主他是什么，我想想他是什么，他走那种，因为他好像很有钱，就是那种住国外，然后不用工作那种，然后我还是个学生，我就有点羡慕你知道吧？我就看完了他的视频看的那一条，他讲的他什么怎么攻，生活怎么样的那种，然后我看了之后，然后他就开始给我推他的穿搭，就一直给我推他的穿搭每一期的穿搭，我就不知道怎么办可能太长了，然后我给他看完了，我觉得这是最重要的原因。
Following users had become more about customizing the feeds than social connection. Participants were less likely to browse the ``Follow'' page on these platforms, as followed users' content was already integrated into their recommendation feeds. P09 stated that it was unnecessary to pay attention to whom she followed because the algorithm would automatically recommend content created by them as well as relevant content:
\begin{quote}
    I've followed 6,000 people. They’re like seeds—I don't care which company produced it as long as they grow into flowers... The algorithm prioritizes content from bloggers I follow, so I will still follow them when necessary, but I’m not interested in remembering their names. (P09)
\end{quote}
% 没必要我灌输6000多个人，我记啥名字？他们只是我所需要的，就像我需要的种子一样，我只要这个种子能种出花就行，我不管他哪个厂生产的，我懒得记，你只要种子好也就。
% Researcher W 27:36
% 你的意思是说您觉得其实不用关注博主，只要是根据您以往的搜索和点击的，模式算法自动会把这些相关的退给你，是这个意思吗？
% Participant 27:52
% 不是不用关注，是不用记住它的名字。我关注还是会关注的，因为它的算法也会考虑到我关注哪几位博主在有的时候手机上不会不是会有个小小小圆圈说你的关注吗？他被推到我眼前的概率不就更高一点，但对于我没有关注，所以我该关注会关注，但是他们具体叫什么我懒得。
Although some participants reported using following and liking to refine recommendation feeds, these actions were more commonly considered natural interactions with the platform to express appreciation, which we categorized as unintentional implicit feedback.

% Many participants mentioned that they often forward posts with friends via direct messages or chat groups if they think their friends would be interested. They sometimes forward these posts to other social media apps they commonly used for daily conversation. This sharing may leads to further discussion. Additionally, participants observed that forwarding and communicating around the content will influence recommendations. For example, P12 mentioned:

% \begin{quote}
%     I frequently interact with two friends, and the three of us have a group chat where they like to share various posts. Some of these posts interest me, while others do not, but I always respond positively out of politeness. I believe that this sharing process affects my homepage recommendations. Since we follow each other, the algorithm might assume that I like their content. Additionally, my positive responses to their posts might signal to the algorithm that I am interested in similar content.
% \end{quote}

Many participants expressed concerns about platforms monitoring their daily conversations both online and offline. One participant (P25) strategically leveraged the perceived surveillance to influence recommendation feeds by \textit{``deliberately talking loudly about certain topics to let the app overhear''} and receive more desirable content. 

\subsection{Purposes for Usage of Personalized Recommendation Platforms}
We categorized users' purposes that motivate users to interact with algorithms into: content consumption, directed information seeking, content creation and promotion, and feed customization. Understanding the purposes provides context for interpreting the feedback and helps tailor recommendation feeds to better meet users' needs. 

We categorized the high-level purposes that motivate users to interact with algorithms into: content consumption, directed information seeking, content creation and promotion, and feed customization.

\subsubsection{Content consumption} 
A common purpose reported by all participants was consuming recommendation content. This referred to browsing the ``Explore'' page in Xiaohongshu undirectedly, or engaging with the continuous stream of short videos in the ``For You'' page of Douyin, Kuaishou, or Bilibili Shorts. Sometimes it became part of a participant’s routine. For instance, P09 used the explore page in Xiaohongshu as a ``library'' to gain creative inspiration related to calligraphy or drawings. When users encountered content they found interesting or useful, they naturally took further actions, such as sharing it with friends, collecting it, or downloading it for later use. Conversely, users' dissatisfaction with the recommendation content, such as perceived low relevance or high homogeneity, would lead to the purpose of feed customization.

\subsubsection{Directed information seeking} 
Another purpose was seeking information more directedly, such as searching for skincare tips, recipes, or travel advice. This often involved validating and cross-checking information, prompting further actions based on the perceived quality and authenticity of the content, such as searching across platforms or consulting friends. Several other participants highlighted the immediacy and relevance of the information on personalized recommendation platforms, noting that they now relied less on search engines for \textit{``everyday inquiries''} (P12). For example, P12 used it to search for available exam centers and compare their conditions, she noted that \textit{``the platform provides timely and useful information, whereas Baidu's (a Chinese search engine) results tend to be outdated or more official.''}

% 我觉得基本上对我来说的话，还是能够给我比较大的信息的，比如说在当地他考试条件如何，他考点是否开放，基本上我觉得都能得到答案。想说某种程度上我觉得现在有一点像百度的那种作用。
% Researcher W 02:24
% 你觉得他跟百度比的话哪个更好，或者说是他们的不同在哪里？
% Participant 02:31
% 我觉得小红书其实比较怎么讲，我觉得算是及时性比较强的，因为它是一个你就感觉是有很多人在回答你的问题，百度的话其实我现在就不太上百度上搜东西了，我感觉可能更多的像搜索这种内容就是比较生活性的，然后百度的问题可能它会比较老，或者说它会比较官方专业性的那种，感觉还是不太一样，像比如生活日常类的东西，我就会在小红书上搜。
% Researcher W 03:03
% 您刚才说会有很多人来回答你指的是搜索行为是吧？搜索了以后会找到一些答案是吗？
% Participant 03:12
% 对，我之前也有发过帖子问一个专门的问题，但那个的话其实可能还是回复量比较少，我觉得这个好像就比较看大数据它会不会给你推送出去或者怎么样，所以说主要还是进行一个搜索的时候，能够得到答案比较多。
% Researcher W 03:31
% 您是说您之前发过一个帖子，那个帖子有有得到多少回复，然后有没有一些回复，你觉得还可以。
% Participant 03:42
% 人家当时新冠的时候就是我自己养的，然后我就想了解一下大概这个症状是不是或者怎么样。但是当时发的话，我看这浏览量是500多，但是回复只有一个，从也等于是对我来说也没有太大的一个用处的，反而是我去搜索一些东西来对应我自己还是比较有用。

\subsubsection{Content creation and promotion}
Several participants shared their experiences of posting content on these platforms, noting that whether a post can achieve widespread reach often hinged on their understanding of the algorithm and how well they could harness it. Committed posters would closely monitor their post traffic and speculate about the underlying mechanisms. P14, for instance, mentioned that the view counts of her posts sometimes caused her anxiety. She attributed the limited traffic to a  lack of comments. P18 mentioned strategically using hashtags and crafting catchy titles to boost the visibility of their posts. By content creation and promotion, participants leaned more about the algorithm. They then not only used this knowledge to increase their posts' visibility as content creators but also curated their own recommendation feeds as content consumers. For example, P12 observed that using specific tags boosted the visibility of her posts. Then, she searched by these tags to obtain more relevant content.

% A few participants also play the role of content creators other than consumer. When they post on these platforms, they often pay attention to the popularity-related data of these posts, such as views and likes. They also speculate about the underlying mechanisms driving this popularity, forming folk theories, and use these theories to achieve more popularity. Some participants also observed that after posting certain content, their recommended feeds would start featuring related topics. P11 reflected this experience:

% \begin{quote}
%     I once posted something about my idol. Although my post did not get much attention, the platform started recommending a lot of popular and related content in my feeds.
% \end{quote}

\subsubsection{Feed customization} 
The need for feed customization arose when users were not satisfied with the personalized recommendation feeds, leading to deliberate actions to shape their content. Our analysis identified four specific goals for feed customization: improving recommendation relevance, increasing content diversity, reducing inappropriate content, and protecting privacy. 

\textit{Improving recommendation relevance} was the most common goal, reported by 23 participants, who proactively took action to increase the presence of more relevant and interested content or reduce uninterested or irrelevant content in their feeds. In particular, participants noticed that while algorithms initially captured their preferences well, they struggled to adapt quickly when those interests changed. This delay in the algorithm's response required users to continuously guide it to keep their feeds relevant and engaging. 

% For example, P12 mentioned customizing the feed when it did not capture her change of interest in time:

% \begin{quote}
%     It (the platform) did correctly guess what I was interested in. But after I finished that exam, I didn’t really need that kind of content anymore, and I even started to resist it. But it kept pushing it to me anyway, so I subconsciously started searching for some unusual things. I felt like I was using my behavior to tell him that I didn’t want to see this anymore. (P12, female)
% \end{quote}
% 他的确是猜我感兴趣，确实猜到了，然后然后但是有的时候也会觉得也完全没必要，满屏都是这个东西，因为我所需要的内容我其实已经知道了，然后就像那段时间我考完那个试之后，我其实就不太需要这方面的内容了，甚至我比较抵触这方面的内容，他依然还是在给我做这些，然后我就有点下意识的想要搜索一些奇特的东西，我觉得能不能相当于是我在用我的行为告诉他，我不想看这也是想给我换别的这种感觉

\textit{Increasing content diversity} was reported by 18 participants who sought to break out of homogeneous content and enrich their feeds with a broader range of topics. Many participants felt tired or annoyed by seeing repetitive content. P15 mentioned she did not want to open Douyin anymore, saying, \textit{``I'm even experiencing browsing fatigue now, often feeling that I’m not exposed to any new ideas.''} Some participants worried that recommendation content too closely aligned with their personal preferences could narrow their perspectives or intensify polarized views. For example, P12 felt the platform was \textit{``deliberately trying to please you, and it wants you to know that it's pleasing you, which is not very smart and can trap you in the information cocoon.''} P28 said \textit{``We used to have our phones follow our minds, but now our minds follow our phones, so the outcome is inevitably becoming narrower and narrower.''} These concerns lead them to proactively increase the content diversity.

\textit{Reducing inappropriate content} was reported by 11 participants who would take action when they encountered content that triggered a strong negative reaction. The goal was primarily driven by a desire to filter out ads or inappropriate and offensive content to maintain a more enjoyable online environment. 

\textit{Protecting privacy} was a less reported feed customization goal. Only two participants mentioned taking action to protect their privacy, despite that many participants expressed repulsion and concern over the platforms' invasion of privacy, such as monitoring posts sharing with friends within the platform, tracking conversations on other social media, and even eavesdropping on offline conversations.

\subsection{Variation in Feedback Types Used Across User Purposes}
After classifying participants' purposes for use of personalized recommendation platforms, we further analyzed the relationship between these purposes and feedback types. We mapped the feedback types to the corresponding purposes through code co-occurrence analysis. The results showed that the feedback types used were highly associated with the purposes. Specifically, intentional implicit feedback and explicit feedback were primarily used for feed customization, with intentional implicit feedback used more towards increasing content diversity and improving recommendation relevance, and explicit feedback for improving recommendation relevance and reducing inappropriate content. Unintentional implicit feedback was most commonly linked to content consumption.

\subsubsection{Feedback types aligning with general purposes} 
The frequency distribution of the three feedback types and four general purposes is presented in \autoref{tab:purposeRelation}. Each participant might have reported multiple feedback behaviors with respective purposes, therefore the total occurrences ($n = 198$) is larger than the number of participants. Unintentional implicit feedback was presented 105 instances among different purposes, followed by intentional implicit feedback (58 instances) and explicit feedback (35 instances).

\begin{table*}[]
\caption{The frequencies of each feedback type---unintentional implicit feedback, intentional implicit feedback, and explicit feedback---across user purposes, including content creation and promotion, feed customization, content consumption, and directed information seeking.}
\label{tab:purposeRelation}
\centering
\resizebox{\textwidth}{!}{%
\begin{tabular}{p{0.3\linewidth}p{0.25\linewidth}p{0.2\linewidth}p{0.15\linewidth}p{0.1\linewidth}}
\toprule
\textbf{Purpose} & \textbf{Unintentional implicit} & \textbf{Intentional implicit} & \textbf{Explicit} & \textbf{Sum} \\
\midrule
Content consumption & 63 & 9 & 1 & 73 \\
Feed customization & 5 & 44 & 31 & 80 \\
Directed information seeking & 25 & 5 & 1 & 31 \\
Content creation / promotion & 12 & 0 & 2 & 14 \\
\midrule
\textbf{Sum} & \textbf{105} & \textbf{58} & \textbf{35} & \textbf{198} \\
\bottomrule
\end{tabular}
}
\end{table*}

Overall, intentional implicit feedback (75.9\%, 44 out of 58 instances) and explicit feedback (88.6\%, 31 out of 35 instances) was predominately utilized for feed customization, while unintentional implicit feedback (60\%, 63 out of 105 instances) was most commonly associated with content consumption.

For content consumption, 63 instances highlighted natural interactions with the platform, such as liking, sharing, or saving posts to express appreciation or for future reference. Participants might also browse the content creator's profile page, view the search prompt, or search for additional information to learn more when something piques their interest. For instance, P18 often expanded the comment section to see the ``Trending searches'' provided by Douyin to quickly find explanations or more related content about the post. All these natural behaviors would then be regarded as unintentional implicit feedback by algorithms to refine recommendation feeds. Although participants performed these behaviors without (reporting) intention of influencing the algorithm, some participants noticed that the algorithm had responded to their behaviors. P18 noticed the platform started to recommend related content after she lingered at someone's profile page.

% Only nine instances opted for intentional implicit feedback by following posts and tracking creators that interested them. 

The purpose of feed customization had 44 instances of intentional implicit feedback, such as ignoring or swiping past a post, clicking on a post to view it, or initiating a search to shape recommendation feeds. For example, P27 mentioned that when customizing his feeds, he would spend more time on videos about snowboarding (new interest) rather than badminton (known interest). Additionally, 31 instances were explicit feedback, such as marking content as ``Not interested'' or blocking it, to actively intervene with the algorithm. %(P06).

For directed information seeking and content creation and promotion, the most common feedback behaviors were searching for information and creating posts respectively. Although in these circumstances, searching and posting actions were not performed deliberately for customizing the feed, the system still interpreted them as feedback and adjusted recommendation content accordingly. P11 noted that after posting about her favorite idol, the platform began recommending more popular and related content in her feed, even though her post itself attracted little attention. Two participants mentioned explicitly using hashtags for content promotion. For example, P05 used ``Note Inspiration''\footnote{A specific feature, just like a trending topic for posts, can be highlighted using hashtags. Posts with these hashtags tend to be clustered around a particular theme, often resulting in higher traffic and engagement.} on Xiaohongshu to promote the visibility of her posts.

%Moreover, there was a notable difference in that more people chose directed information seeking compared to self-posting on short video platforms, which were initially designed for social interaction with friends.

These findings demonstrated observable differences in feedback types when the user's intent was to customize feeds compared to when they engaged in activities like content creation or consumption. While feed customization primarily relied on intentional implicit and explicit feedback, other purposes are more associated with unintentional implicit feedback. 

\subsubsection{Feedback types driven by specific feed customization purposes}

\begin{table*}[]
\centering
\small
\caption{The frequencies of intentional implicit feedback and explicit feedback across four specific feed customization purposes, including increasing content diversity, improving recommendation relevance, reducing inappropriate content, and protecting privacy.}
%Unintentional implicit feedback is not included in this table because there were limited instances of its use for algorithm customization purpose.
\label{tab:feedback_algopurpose}
\resizebox{\textwidth}{!}{%
\begin{tabular}{p{0.3\linewidth}p{0.3\linewidth}p{0.2\linewidth}p{0.1\linewidth}}
\toprule
\textbf{Feed customization purpose} & \textbf{Intentional implicit feedback} & \textbf{Explicit feedback} & \textbf{Sum} \\ \midrule
Improving recommendation relevance & 18 & 14 & 32\\
Increasing content diversity & 24 & 5 & 29\\
Reducing inappropriate content & 2 & 10 & 12\\
Protecting privacy & 0 & 2 &  2\\ \midrule
\textbf{Sum} & \textbf{44} & \textbf{31 }& \textbf{75}\\
\bottomrule
\end{tabular}%
}
\vspace*{-8pt}
\end{table*}

% Most participants noted that the algorithms are generally effective at capturing their intentions and responding to their feedback, although it may require multiple interactions to observe the changes. However, this same accuracy in recommendations can also raise concerns among some users to protect their privacy or avoid the effects of an information cocoon. There are instances when participants decide they no longer want to continue with algorithm training and choose to stop using the platform. P15: 
% \begin{quote}
%     I feel that most of the viewpoints are homogeneous, and it's rare to encounter differing perspectives. I'm even experiencing browsing fatigue now—often feeling that I’m not exposed to any new ideas, and the content I see is incredibly dull. This has led me to either close the app or not open TikTok at all. (P15, female)
% \end{quote}

As \autoref{tab:feedback_algopurpose} depicts, for users aiming to improve recommendation relevance, both intentional implicit feedback (18 instances) and explicit feedback (14 instances) were frequently used.
% , while unintentional implicit feedback was rarely observed; therefore, it will not be discussed here. 
Ignoring or swiping past a post was the most used intentional implicit feedback, while marking as ``Not interested'' was the most common explicit feedback. P27 asserted that swiping away the content was already effective in signaling disinterest, eliminating the need for explicit marking. P11 explained that her choice between these feedback methods depended on the context and the degree of dislike. She would only use the marking as ``Not interested'' option when she \textit{``deeply detested''} the recommendation content. 
\begin{quote}
    I hadn't searched for ``The Wandering Earth'' before, but it still recommended it to me. This might be because ``The Wandering Earth'' is very popular lately, so the system tries to see if I'm interested. If I don't pay attention, they'll likely disappear. So, with these exploratory recommendations, unless I deeply detest them, I usually just swipe them away to get some new content. (P11)
\end{quote}
% 我觉得他可能会根据一些现在实时的热点推给我，比如说我之前说我没有搜索过流浪地球，但是他却推给我了，这样的情况可能就是因为流浪地球最近很火，所以他就尝试着推给我，看我有没有看过，但一般这种推送都非常的少，如果我不去注意的话，它可能就过去了，所以一般有这种试探性的推送，除非我真的对它深恶痛绝，我会点个不感兴趣，一般我就直接去刷新一下，换一批那种感觉

For detailed associations between specific feed customization purposes and feedback types, as illustrated in \autoref{tab:feedback_algopurpose}, when participants sought to improve content diversity, 24 out of 29 instances used intentional implicit feedback, with only five opting for explicit feedback. The primary implicit feedback behavior was initiating a new search. The participants discovered that searching certain topics increased the weight of related content in recommendation feeds, thus they strategically searched for new topics. As P06 speculated: \textit{``the system might update your user profile or data based on that search and then start pushing related content that aligns with your new interests.''}

In contrast, 10 out of 12 instances chose explicit feedback to reduce inappropriate content. The explicit feedback was perceived as quicker and more direct to address inappropriate content. Participants tended to mark content as ``Not interested'' when they found it low-quality, full of ads, or disturbing (P09 and P16). In cases of offensive personal attacks or strong disagreement with a creator's opinion, users were more likely to report or block. 
% P26 mentioned, ``I strongly disagreed with his (the blogger's) opinion, so I blocked him,'' and P34 commented:
% \begin{quote}
%     The internet is full of all kinds of weird things, and indeed, there were some rather odd users. Often, especially when someone behaved rudely, there was a tendency for me to block or delete them. (P34, female)
% \end{quote}

% P27 believed that this feature allowed users to control the content being recommended to them:
% \begin{quote}
%     Making good use of the search function was important. You needed to divide yourself into two modes: one for browsing aimlessly out of boredom, and another with a purpose. It was like using Bilibili; you needed to use it in two ways. It had both the features of Bilibili and Douyin; it just depended on how you switched between them. (P27, male)
% \end{quote}

Only two participants mentioned their attempt to protect privacy by explicitly disabling the personalization option. P01, concerned about the platform's data collection and user profiling practices, decided to turn off personalization to prevent Xiaohongshu from using his interactions to build a detailed user profile. P14 said that she tried disabling personalization to protect privacy, however, without knowing \textit{``whether it really worked.''} 
%These users did not receive clear feedback on whether their privacy was effectively protected.%

The results underscored how users' feed customization goals, along with their understanding of algorithms, drove their choice of feedback type for the personalized recommendation platforms. Overall, participants relied more on intentional implicit feedback to increase content diversity, whereas explicit feedback is more preferred for reducing inappropriate content. Participants aiming to improve recommendation relevance employed a mix of both implicit and explicit feedback. Meanwhile, participants concerned with protecting privacy took a more passive approach, seldom engaging with feedback mechanisms, in contrast to the active use of intentional implicit feedback by those aiming to increase content diversity. 

\section{Discussion}
\subsection{Revisiting the Explicit-Implicit Dichotomy Through Intention}
% 1. Why our findings of the ``intentional implicit feedback'' are important. We can talk about the benefits: identifying the intentional implicit feedback can (1) help with users' understanding and control of the algo and achieve more desirable personal recommendation, (2) reduce user unintended or unexpected outcome (false positive error), and (3) free up users' cognition for other tasks. 
% We can also talk about the problems of the explicit–implicit dichotomy...

The explicit-implicit feedback dichotomy was largely framed from the system's perspective, as Jannach et al. framed it---implicit feedback referred to interactions from which we (the system side) can indirectly infer user preferences ~\cite{jannach2018recommending}. The implicit feedback was supposed to be user's natural interactions with the platform. However, from the user’s perspective, they are not only aware of, but also intentionally, proactively act as feedback inputs to influence the recommendation feeds. For instance, in our findings, when participants ``dwell on'' or ``swipe past'' a post, they might consider it as an explicit expression of their preferences. Yet, the system might take it as implicit feedback and fail to understand or capture it. This highlights a fundamental challenge: feedback categorized as implicit from the system's point of view may be deliberately used as explicit by users.

This misalignment between how users and the system perceive feedback can lead to inaccuracies or failures in capturing user intent. A specific problem is that implicit feedback is often only interpreted as positive feedback~\cite{jawaheer2014modeling,hu2008collaborative}, whereas our findings revealed that users also intentionally employed implicit feedback behaviors to communicate negative feedback, such as ignoring or swiping past a post ($n = 21$), stop using the platform or switch platforms ($n = 12$), and refresh the feed ($n = 2$). Neglecting on interpreting negative implicit feedback could distort the user profile~\cite{hu2008collaborative}. Thus, incorporating both positive and negative implicit feedback is essential for more accurate recommendation. Another related problem is that the platform' focus on implicit positive feedback may lead to instances of false positive interpretation. For instance, several participants mentioned they searched for information as a one-time task, but the platform over interpreted this as a preference, resulting in their feed being flooded with similar content that was no longer relevant or needed. This suggests that implicit feedback should be considered through more nuanced distinctions, such as positive versus negative, intentional versus incidental~\cite{dix2002beyond}, and reactive versus proactive behaviors~\cite{ju2008range}. Particularly, defining explicit–implicit distinction through intention can add greater precision to the interpretation of user feedback~\cite{serim2019explicating}. 
% implicit feedback has traditionally been treated as positive, and the lack of attention to negative signals can lead to misrepresenting user preferences ~\cite{hu2008collaborative}. However, our participants frequently used implicit feedback to express negative signals, with actions like ignoring or swiping past posts being the most common.

%Therefore, the addition of intentional implicit feedback to the implicit-explicit dichotomy in our paper can help platforms better distill users' intention, addressing the above issues to improve the accuracy in predicting user preferences, achieving more desired personalized recommendations.Comparing to unintentional implicit feedback, 
In addition, acknowledging and responding to the intentional implicit feedback can afford users more understanding of the algorithm’s learning process. Users will feel that their interactions are shaping the recommendation feeds, thus foster a greater sense of self-causality and increase user agency~\cite{feng2024mapping}. Comparing to explicit feedback, intentional implicit feedback also frees up users’ cognitive resources. Research has shown that explicit feedback often imposes a higher cognitive load, as users must actively engage with the system ~\cite{kelly2003implicit,gadanho2007addressing}. In contrast, allowing users to guide recommendations through unobtrusive, yet intentional, behaviors can make the interaction more seamless and less mentally taxing. This aligns with findings from previous studies that highlight the importance of minimizing user effort in feedback mechanisms.

\subsection{Aligning Feedback Types with Users' Purposes}
% Discuss why users chose to use different feedback types for different purposes. The nature and characteristics of the feedbacks.
The study identified four purposes in participants' using of personalized recommendation platforms: content consumption, directed information seeking, content creation and promotion, and feed customization. The first three purposes echo previous literature, which has discussed directed and undirected consumption~\cite{feng2024mapping}, as well as self-presentation ~\cite{devito2017algorithms}. We identified feed customization as a purpose that arises when users are dissatisfied with the recommendation contents, which encompasses four specific goals. 

Users choose different feedback types based on their purposes. Explicit feedback (75.9\% instances) is primarily used for feed customization. As shown in~\autoref{tab:feedbackType}, all explicit feedback behaviors were supported by platform features, providing cues to helps users understand the potential outcomes of their actions. Such clarity makes explicit feedback the preferred choice when users seek quick corrections to their recommendation feeds. For instance, our findings indicated that most users opted for explicit feedback, particularly negative explicit feedback, such as marking content as ``Not interested,'' blocking, or reporting, when they had a strong desire to remove inappropriate or irrelevant content from their feeds. Previous studies found that users were reluctant to provide explicit feedback in complex tasks like online shopping but were more willing to do so in media and entertainment domains~\cite{white2005study,jawaheer2014modeling}.
This aligns with our findings that, on personalized recommendation platforms, users did not show a strong preference between explicit (31 instances) and implicit feedback (44 instances) during feed customization.

On the other hand, intentional implicit feedback (88.6\%) was also mostly used for feed customization, but often when users' intent is not particularly strong or urgent---either to increase content diversity or improve recommendation relevance. Intentional implicit feedback allows users to steer the algorithm towards their evolving preferences without overtly signaling their intent or interrupting the flow of content consumption. The choice among various implicit feedback behaviors is usually based on their continuously updated folk theories~\cite{devito2017algorithms}. Since users are not fully certain how the algorithm will interpret their actions ~\cite{eslami2015always}, they tend to experiment and make repeated attempts when necessary. Despite these efforts, our participants sometimes observed minimal or delayed changes in the recommendation feeds, due to the less obvious and plausible nature of implicit feedback~\cite{jannach2018recommending}. This then may then lead to more passive behaviors like stop using the platforms.

Moreover, the design and functionality of the platforms~\cite{liu2010personalized}, such as Douyin’s swipe interaction versus Xiaohongshu’s click-and-select feed layout, also mediate users' specific feedback behaviors. Participants on Xiaohongshu, for example, found selecting and clicking more efficient for improving recommendation relevance, comparing to using the swipe function in Douyin. Therefore, platform design should support a wider range of intentional implicit feedback by providing users with clearer traits of how their actions influence recommendation feeds and more intuitive ways to interact with the platform.

\subsection{Design Implications}
% 1. supporting intentional implicit feedback: (1) provide features to support or prompts, but also remain unobtrusive for natural flow. (2) support negative feedback.Most used implicit and explicit are both negative feedback.
% (3)  The need for users to repeatedly mark content as ``Not interested'' suggests that the algorithm may not be efficiently learning from initial feedback. Enhancing the algorithm's ability to quickly adapt to user preferences after receiving negative feedback could improve the user experience by reducing the frequency of unwanted recommendations. but also not too sensitive to reduce false positive 
% 2. Incorporate user goals. such as balance between relevance and diversity
% 3. transparency in data collected and protect privacy 
The discussion leads to the following design implications for the personalized recommendation platforms to effectively support and respond to user feedback driven by various purposes.

\subsubsection{Supporting Intentional Implicit Feedback Through Recognition and Visualization} 
First, platforms need to support intentional implicit feedback. When users intentionally provide implicit feedback, they expect the algorithm to infer their interest from their ``natural interactions''~\cite{kelly2003implicit} and respond seamlessly. This differs from explicit feedback, which is given through direct user input that interrupts users' natural flow. However, when the subtle feedback goes unnoticed or yields little immediate effect, users may feel a loss of control. Therefore, while most implicit feedback is not supported by platform features (shown in~\autoref{tab:feedbackType}), it's important to inform users that their feedback has been recognized and will influence future recommendation contents. One potential solution is to implement non-intrusive visual cues, such as small icons or pop-ups, that confirm feedback has been registered. For instance, after a user click similar posts several times or search for a topic, a pop-up could confirm that the platform has recognized this interest and will increase similar content recommendations. Another option is to provide progress indicators that allow users to see how their behavior is affecting the feeds over time. This could include a dashboard showing how many times they had skipped certain content and how it has adjusted their recommendation feeds. This can increase transparency and help users understand how the algorithm is learning from their feedback and responding accordingly. 

Additionally, platforms need to improve their handling of negative implicit feedback. The challenge with interpreting negative implicit feedback lies in its subtlety. Unlike explicit feedback like marking as not interested, implicit negative signals are often less direct and harder to capture. To address this issue, platforms need to adopt more sophisticated approaches to detect and validate implicit negative feedback. While some platforms have already implemented ``show less'' features that allow users to indicate their content preferences (such as ~\cite{meta_new_2022}). These features are typically static and presented with all content. What's needed is a more dynamic approach that can capture and respond to users' behavioral patterns in real-time. One potential strategy is explicit user confirmation. For example, when a user repeatedly swipes past posts or consistently ignores content, the system could proactively detect that the user is not interested in this topic and prompt the user with an option like, ``Want to see less of this content?'' to confirm their intent. Such dynamic approaches can capture users' intentional implicit feedback behaviors more effectively and would not cause a significant disruption to users' normal usage.
% acknolwege current available design(https://ai.meta.com/blog/facebook-feed-improvements-ai-show-more-less/#:~:text=The%20challenge%20with%20training%20these,they%20would%20like%20to%20see)

\subsubsection{Designing for Purpose-Oriented Feedback}
Second, we suggest designing for purpose-oriented feedback. Our findings revealed close association between users' purposes and their choice of feedback types. This underscores the importance of incorporating users' diverse and evolving purposes in the design of personalized recommendation platforms in addition to focusing on their behaviors~\cite{liang2023enabling, nazari2022choice}. To address this, platforms can integrate context-aware learning models to infer user purposes. For example, during directed information seeking, algorithm can prioritize relevance and rely more on implicit feedback~\cite{white2005study}; in content consumption, the algorithm can also consider content diversity, as many participants were concerned with information cocoon~\cite{li2022exploratory}. Meanwhile, the platforms can also provide customization options, such as a slider, for users to adjust the balance between relevance and diversity in their recommendation contents. It should be mention that these setting options, such as setting up interested channels, should be made more discoverable, as some participants did not notice the feature~\cite{liu2024train}.

\subsubsection{Transparency in Feedback Data Collection and Protecting Privacy}
Quite a few participants expressed concerns about their privacy, noting that platforms seemed to be monitoring their conversations, chats with friends, and even activities on other platforms. Despite these concerns, only two participants took active steps to protect their privacy by disabling personalization. However, after noticing that the recommendation quality significantly deteriorated, they eventually re-enabled it. This suggests that users have no effective means to combat platforms' invasion of privacy. To address these concerns, platforms should be transparent about the types of data being collected, including behaviors within the platform, cross-platform tracking, and even offline monitoring. Platforms should allow users to customize their privacy settings, giving them the ability to select which types of data they are comfortable sharing for algorithm personalization. This could include options to opt out of specific data types, while still maintaining a personalized recommendation system.

Interestingly, as we mentioned in the results, one participant mentioned deliberately speaking aloud to trigger the platform's monitoring and influence recommendation feeds. This behavior highlights a potential design opportunity: platforms could introduce voice-based feedback to explicitly capture users' preferences, rather than continuously monitoring their voices. By offering users the ability to verbally express their content needs, platforms could create a more transparent and interactive method for users to directly influence recommendation feeds while respecting their privacy.

% 4. What will be the challenges of designing for ``intentional implicit feedback''? For example, ``constructive validity''~\cite{serim2019explicating}---the extent about how intentional implicit feedback translates into design in our case. 
% Other challenges can be users' literacy to use these designs, users' fluid needs, and delayed system reactions. According to Serim and Jacucci~\cite{serim2019explicating}, users' attitudes can be fluid and influenced by situational factors, which poses uncertainties to determining implicitness.
\subsection{Limitations and Future Work}
This study has several limitations that should be noted. First, the recruitment method relied on pre-screening surveys and snowball sampling. This might have introduced selection bias, favoring more engaged platform users and educated users. Second, the feedback behaviors reported by users were neither comprehensive nor reflective of the ground truth. Specifically, the list of unintentional implicit feedback behaviors in ~\autoref{appendix:unintentional} includes only interactions mentioned by users and categorized as implicit feedback in previous literature~\cite{jannach2018recommending}. We do not have access to the complete set of behaviors that platforms use as implicit feedback. 

% \subsection{Future Work}
% 研究什么样的人会用intentional里的所有behavior，和他们的demo characteristics，以及为何选择specific behavior
Our findings shed light on how users leveraged different feedback mechanisms to fulfill their purposes on personalized recommendation platforms, but further investigation is needed to validate and expand upon these insights. One potential future work is to conduct a large-scale survey to confirm the feedback behaviors and patterns identified in our study. Although our interviews touched upon users' motivations for selecting different types of feedback in various contexts and purpose-oriented situations, a survey study would allow us to explore users' rationales when choosing different feedback types in different contexts. Another important direction for future research is to examine how different types of feedback—explicit feedback, unintentional implicit feedback, and intentional implicit feedback—impact user experience and satisfaction with recommendation systems. This could include how each feedback type influences content quality, user engagement, and perceived control over the recommendation system. Additionally, future research could examine the relationships between users' feedback choices and factors such as demographics and algorithm literacy. Understanding how these factors influence users' engagement with feedback mechanisms could support the design of more personalized and effective feedback mechanisms.
\section{Conclusion}

We conducted semi-structured interviews with 34 active users on platforms like Xiaohongshu and Douyin to explore how users employ diverse feedback mechanisms to influence recommendation feeds towards specific purposes. We categorized various user feedback behaviors into three types: explicit, intentional implicit, and unintentional implicit feedback. We also found that users' choices of feedback types are closely aligned to their purposes. Explicit feedback was primarily used for feed customization goals like reducing inappropriate content and improving recommendation relevance, while intentional implicit feedback emerged as crucial for feed customization to increase content diversity and improve recommendation relevance. Unintentional implicit feedback was most commonly linked to content consumption. The study introduced the intention dimension into the traditional explicit-implicit feedback dichotomy and suggested that personalized recommendation platforms should better support transparent intentional implicit feedback and purpose-oriented feedback design.

\begin{acks}
We thank the participants for sharing their experiences and insights. We would also like to express our appreciation to Zitong Huang and Yaqi Zhang for their assistance in collecting and analyzing the interview data. Additionally, we extend our thanks to Lu Xian, Zefeng Ben Zhang, and all reviewers for their thoughtful feedback, which greatly contributed to refining this work. This research was supported by the NSFC Grant No.72174014 and funding from the Inequality in America Initiative at Harvard University. 
\end{acks}

\balance
\bibliographystyle{ACM-Reference-Format}
\bibliography{0-references, references_poster}


\begin{thebibliography}{86}

%%% the Web address.

\ifx \showCODEN    \undefined \def \showCODEN     #1{\unskip}     \fi
\ifx \showDOI      \undefined \def \showDOI       #1{#1}\fi
\ifx \showISBNx    \undefined \def \showISBNx     #1{\unskip}     \fi
\ifx \showISBNxiii \undefined \def \showISBNxiii  #1{\unskip}     \fi
\ifx \showISSN     \undefined \def \showISSN      #1{\unskip}     \fi
\ifx \showLCCN     \undefined \def \showLCCN      #1{\unskip}     \fi
\ifx \shownote     \undefined \def \shownote      #1{#1}          \fi
\ifx \showarticletitle \undefined \def \showarticletitle #1{#1}   \fi
\ifx \showURL      \undefined \def \showURL       {\relax}        \fi
% invisible to TeX
\providecommand\bibfield[2]{#2}
\providecommand\bibinfo[2]{#2}
\providecommand\natexlab[1]{#1}
\providecommand\showeprint[2][]{arXiv:#2}

\bibitem[Adomavicius and Tuzhilin(2005)]%
        {adomavicius2005toward}
\bibfield{author}{\bibinfo{person}{Gediminas Adomavicius} {and} \bibinfo{person}{Alexander Tuzhilin}.} \bibinfo{year}{2005}\natexlab{}.
\newblock \showarticletitle{Toward the next generation of recommender systems: A survey of the state-of-the-art and possible extensions}.
\newblock \bibinfo{journal}{\emph{IEEE transactions on knowledge and data engineering}} \bibinfo{volume}{17}, \bibinfo{number}{6} (\bibinfo{year}{2005}), \bibinfo{pages}{734--749}.
\newblock


\bibitem[Bambauer and Zarsky(2018)]%
        {bambauer2018algorithm}
\bibfield{author}{\bibinfo{person}{Jane~R Bambauer} {and} \bibinfo{person}{Tal Zarsky}.} \bibinfo{year}{2018}\natexlab{}.
\newblock \showarticletitle{The algorithm game}.
\newblock \bibinfo{journal}{\emph{Notre Dame L. Rev.}}  \bibinfo{volume}{94} (\bibinfo{year}{2018}), \bibinfo{pages}{1}.
\newblock


\bibitem[Bishop(2018)]%
        {bishop2018anxiety}
\bibfield{author}{\bibinfo{person}{Sophie Bishop}.} \bibinfo{year}{2018}\natexlab{}.
\newblock \showarticletitle{Anxiety, panic and self-optimization: Inequalities and the YouTube algorithm}.
\newblock \bibinfo{journal}{\emph{Convergence}} \bibinfo{volume}{24}, \bibinfo{number}{1} (\bibinfo{year}{2018}), \bibinfo{pages}{69--84}.
\newblock


\bibitem[Bishop(2019)]%
        {bishop2019managing}
\bibfield{author}{\bibinfo{person}{Sophie Bishop}.} \bibinfo{year}{2019}\natexlab{}.
\newblock \showarticletitle{Managing visibility on YouTube through algorithmic gossip}.
\newblock \bibinfo{journal}{\emph{New media \& society}} \bibinfo{volume}{21}, \bibinfo{number}{11-12} (\bibinfo{year}{2019}), \bibinfo{pages}{2589--2606}.
\newblock


\bibitem[Braun et~al\mbox{.}(2019)]%
        {braun_thematic_2019}
\bibfield{author}{\bibinfo{person}{Virginia Braun}, \bibinfo{person}{Victoria Clarke}, \bibinfo{person}{Nikki Hayfield}, {and} \bibinfo{person}{Gareth Terry}.} \bibinfo{year}{2019}\natexlab{}.
\newblock \showarticletitle{Thematic Analysis}.
\newblock In \bibinfo{booktitle}{\emph{Handbook of Research Methods in Health Social Sciences}}, \bibfield{editor}{\bibinfo{person}{Pranee Liamputtong}} (Ed.). \bibinfo{publisher}{Springer, Singapore}.
\newblock
\showISBNx{978-981-10-5251-4}
\urldef\tempurl%
\url{https://doi.org/10.1007/978-981-10-5251-4_103}
\showDOI{\tempurl}


\bibitem[Brulé(2020)]%
        {emeline_brule_thematic_2020}
\bibfield{author}{\bibinfo{person}{Emeline Brulé}.} \bibinfo{year}{2020}\natexlab{}.
\newblock \bibinfo{booktitle}{\emph{Thematic analysis in {HCI}}}.
\newblock
\urldef\tempurl%
\url{https://sociodesign.hypotheses.org/555}
\showURL{%
\tempurl}
\newblock
\shownote{Design and society}.


\bibitem[Bucher(2016)]%
        {bucher2016algorithmic}
\bibfield{author}{\bibinfo{person}{Taina Bucher}.} \bibinfo{year}{2016}\natexlab{}.
\newblock \showarticletitle{The algorithmic imaginary: exploring the ordinary affects of Facebook algorithms}.
\newblock \bibinfo{journal}{\emph{INFORMATION, COMMUNICATION \& SOCIETY}} \bibinfo{volume}{20}, \bibinfo{number}{1} (\bibinfo{year}{2016}), \bibinfo{pages}{30--44}.
\newblock


\bibitem[Bucher(2018)]%
        {bucher2018cleavage}
\bibfield{author}{\bibinfo{person}{Taina Bucher}.} \bibinfo{year}{2018}\natexlab{}.
\newblock \showarticletitle{Cleavage-Control: Stories of Algorithmic Culture and Power in the Case of the YouTube" Reply Girls"}.
\newblock In \bibinfo{booktitle}{\emph{A networked self and platforms, stories, connections}}. \bibinfo{publisher}{Routledge}, \bibinfo{pages}{125--143}.
\newblock


\bibitem[Burrell et~al\mbox{.}(2019)]%
        {burrell2019users}
\bibfield{author}{\bibinfo{person}{Jenna Burrell}, \bibinfo{person}{Zoe Kahn}, \bibinfo{person}{Anne Jonas}, {and} \bibinfo{person}{Daniel Griffin}.} \bibinfo{year}{2019}\natexlab{}.
\newblock \showarticletitle{When users control the algorithms: Values expressed in practices on twitter}.
\newblock \bibinfo{journal}{\emph{Proceedings of the ACM on human-computer interaction}} \bibinfo{volume}{3}, \bibinfo{number}{CSCW} (\bibinfo{year}{2019}), \bibinfo{pages}{1--20}.
\newblock


\bibitem[Cen et~al\mbox{.}(2024)]%
        {Cen_Ilyas_Allen_Li_Madry_2024}
\bibfield{author}{\bibinfo{person}{Sarah~H. Cen}, \bibinfo{person}{Andrew Ilyas}, \bibinfo{person}{Jennifer Allen}, \bibinfo{person}{Hannah Li}, {and} \bibinfo{person}{Aleksander Madry}.} \bibinfo{year}{2024}\natexlab{}.
\newblock \showarticletitle{Measuring Strategization in Recommendation: Users Adapt Their Behavior to Shape Future Content}.
\newblock  \bibinfo{number}{arXiv:2405.05596} (\bibinfo{date}{May} \bibinfo{year}{2024}).
\newblock
\urldef\tempurl%
\url{http://arxiv.org/abs/2405.05596}
\showURL{%
\tempurl}
\newblock
\shownote{arXiv:2405.05596 [cs, stat]}.


\bibitem[Chen et~al\mbox{.}(2019)]%
        {Chen2019}
\bibfield{author}{\bibinfo{person}{Zhuang Chen}, \bibinfo{person}{Qian He}, \bibinfo{person}{Zhifei Mao}, \bibinfo{person}{Hwei-Ming Chung}, {and} \bibinfo{person}{Sabita Maharjan}.} \bibinfo{year}{2019}\natexlab{}.
\newblock \showarticletitle{A study on the characteristics of douyin short videos and implications for edge caching}. In \bibinfo{booktitle}{\emph{Proceedings of the ACM Turing Celebration Conference - China}} (Chengdu, China) \emph{(\bibinfo{series}{ACM TURC '19})}. \bibinfo{publisher}{Association for Computing Machinery}, \bibinfo{address}{New York, NY, USA}, Article \bibinfo{articleno}{13}, \bibinfo{numpages}{6}~pages.
\newblock
\showISBNx{9781450371582}
\urldef\tempurl%
\url{https://doi.org/10.1145/3321408.3323082}
\showDOI{\tempurl}


\bibitem[Cheung and To(2022)]%
        {cheung2022influences}
\bibfield{author}{\bibinfo{person}{Millissa~FY Cheung} {and} \bibinfo{person}{WM To}.} \bibinfo{year}{2022}\natexlab{}.
\newblock \showarticletitle{What influences people to click ‘like’on posts of branded content?}
\newblock \bibinfo{journal}{\emph{Journal of Strategic Marketing}} (\bibinfo{year}{2022}), \bibinfo{pages}{1--23}.
\newblock


\bibitem[Cotter(2019)]%
        {cotter2019playing}
\bibfield{author}{\bibinfo{person}{Kelley Cotter}.} \bibinfo{year}{2019}\natexlab{}.
\newblock \showarticletitle{Playing the visibility game: How digital influencers and algorithms negotiate influence on Instagram}.
\newblock \bibinfo{journal}{\emph{New media \& society}} \bibinfo{volume}{21}, \bibinfo{number}{4} (\bibinfo{year}{2019}), \bibinfo{pages}{895--913}.
\newblock


\bibitem[DeVito et~al\mbox{.}(2018)]%
        {devito2018people}
\bibfield{author}{\bibinfo{person}{Michael~Ann DeVito}, \bibinfo{person}{Jeremy Birnholtz}, \bibinfo{person}{Jeffery~T Hancock}, \bibinfo{person}{Megan French}, {and} \bibinfo{person}{Sunny Liu}.} \bibinfo{year}{2018}\natexlab{}.
\newblock \showarticletitle{How people form folk theories of social media feeds and what it means for how we study self-presentation}. In \bibinfo{booktitle}{\emph{Proceedings of the 2018 CHI conference on human factors in computing systems}}. \bibinfo{pages}{1--12}.
\newblock


\bibitem[DeVito et~al\mbox{.}(2017)]%
        {devito2017algorithms}
\bibfield{author}{\bibinfo{person}{Michael~Ann DeVito}, \bibinfo{person}{Darren Gergle}, {and} \bibinfo{person}{Jeremy Birnholtz}.} \bibinfo{year}{2017}\natexlab{}.
\newblock \showarticletitle{"Algorithms ruin everything" \#RIPTwitter, Folk Theories, and Resistance to Algorithmic Change in Social Media}. In \bibinfo{booktitle}{\emph{Proceedings of the 2017 CHI conference on human factors in computing systems}}. \bibinfo{pages}{3163--3174}.
\newblock


\bibitem[DeVos et~al\mbox{.}(2022)]%
        {devos2022toward}
\bibfield{author}{\bibinfo{person}{Alicia DeVos}, \bibinfo{person}{Aditi Dhabalia}, \bibinfo{person}{Hong Shen}, \bibinfo{person}{Kenneth Holstein}, {and} \bibinfo{person}{Motahhare Eslami}.} \bibinfo{year}{2022}\natexlab{}.
\newblock \showarticletitle{Toward User-Driven Algorithm Auditing: Investigating users’ strategies for uncovering harmful algorithmic behavior}. In \bibinfo{booktitle}{\emph{Proceedings of the 2022 CHI conference on human factors in computing systems}}. \bibinfo{pages}{1--19}.
\newblock


\bibitem[Dix(2002)]%
        {dix2002beyond}
\bibfield{author}{\bibinfo{person}{Alan Dix}.} \bibinfo{year}{2002}\natexlab{}.
\newblock \showarticletitle{Beyond intention-pushing boundaries with incidental interaction}. In \bibinfo{booktitle}{\emph{Proceedings of Building Bridges: Interdisciplinary Context-Sensitive Computing, Glasgow University}}, Vol.~\bibinfo{volume}{9}. \bibinfo{pages}{1--6}.
\newblock


\bibitem[Ellison et~al\mbox{.}(2020)]%
        {ellison2020we}
\bibfield{author}{\bibinfo{person}{Nicole~B Ellison}, \bibinfo{person}{Penny Tri{\^e}u}, \bibinfo{person}{Sarita Schoenebeck}, \bibinfo{person}{Robin Brewer}, {and} \bibinfo{person}{Aarti Israni}.} \bibinfo{year}{2020}\natexlab{}.
\newblock \showarticletitle{Why we don’t click: Interrogating the relationship between viewing and clicking in social media contexts by exploring the “non-click”}.
\newblock \bibinfo{journal}{\emph{Journal of Computer-Mediated Communication}} \bibinfo{volume}{25}, \bibinfo{number}{6} (\bibinfo{year}{2020}), \bibinfo{pages}{402--426}.
\newblock


\bibitem[Eriksson~Krutr{\"o}k(2021)]%
        {eriksson2021algorithmic}
\bibfield{author}{\bibinfo{person}{Moa Eriksson~Krutr{\"o}k}.} \bibinfo{year}{2021}\natexlab{}.
\newblock \showarticletitle{Algorithmic closeness in mourning: Vernaculars of the hashtag \# grief on TikTok}.
\newblock \bibinfo{journal}{\emph{Social Media+ Society}} \bibinfo{volume}{7}, \bibinfo{number}{3} (\bibinfo{year}{2021}), \bibinfo{pages}{20563051211042396}.
\newblock


\bibitem[Eslami et~al\mbox{.}(2016)]%
        {eslami2016first}
\bibfield{author}{\bibinfo{person}{Motahhare Eslami}, \bibinfo{person}{Karrie Karahalios}, \bibinfo{person}{Christian Sandvig}, \bibinfo{person}{Kristen Vaccaro}, \bibinfo{person}{Aimee Rickman}, \bibinfo{person}{Kevin Hamilton}, {and} \bibinfo{person}{Alex Kirlik}.} \bibinfo{year}{2016}\natexlab{}.
\newblock \showarticletitle{First I "like" it, then I hide it: Folk Theories of Social Feeds}. In \bibinfo{booktitle}{\emph{Proceedings of the 2016 CHI conference on human factors in computing systems}}. \bibinfo{pages}{2371--2382}.
\newblock


\bibitem[Eslami et~al\mbox{.}(2015)]%
        {eslami2015always}
\bibfield{author}{\bibinfo{person}{Motahhare Eslami}, \bibinfo{person}{Aimee Rickman}, \bibinfo{person}{Kristen Vaccaro}, \bibinfo{person}{Amirhossein Aleyasen}, \bibinfo{person}{Andy Vuong}, \bibinfo{person}{Karrie Karahalios}, \bibinfo{person}{Kevin Hamilton}, {and} \bibinfo{person}{Christian Sandvig}.} \bibinfo{year}{2015}\natexlab{}.
\newblock \showarticletitle{"I always assumed that I wasn't really that close to [her]" Reasoning about Invisible Algorithms in News Feeds}. In \bibinfo{booktitle}{\emph{Proceedings of the 33rd annual ACM conference on human factors in computing systems}}. \bibinfo{pages}{153--162}.
\newblock


\bibitem[Feng et~al\mbox{.}(2024)]%
        {feng2024mapping}
\bibfield{author}{\bibinfo{person}{K~J~Kevin Feng}, \bibinfo{person}{Xander Koo}, \bibinfo{person}{Lawrence Tan}, \bibinfo{person}{Amy Bruckman}, \bibinfo{person}{David~W McDonald}, {and} \bibinfo{person}{Amy~X Zhang}.} \bibinfo{year}{2024}\natexlab{}.
\newblock \showarticletitle{Mapping the Design Space of Teachable Social Media Feed Experiences}. In \bibinfo{booktitle}{\emph{Proceedings of the CHI Conference on Human Factors in Computing Systems}}. \bibinfo{pages}{1--20}.
\newblock


\bibitem[{Flow Asia}(2023)]%
        {flow2023demographic}
\bibfield{author}{\bibinfo{person}{{Flow Asia}}.} \bibinfo{year}{2023}\natexlab{}.
\newblock \bibinfo{title}{Demographic Overview of Chinese Video Platforms}.
\newblock
\newblock
\urldef\tempurl%
\url{https://www.flow.asia/insight/demographic-chinese-video-platforms/}
\showURL{%
\tempurl}
\newblock
\shownote{Accessed: September 1, 2024}.


\bibitem[Gadanho and Lhuillier(2007)]%
        {gadanho2007addressing}
\bibfield{author}{\bibinfo{person}{Sandra~Clara Gadanho} {and} \bibinfo{person}{Nicolas Lhuillier}.} \bibinfo{year}{2007}\natexlab{}.
\newblock \showarticletitle{Addressing uncertainty in implicit preferences}. In \bibinfo{booktitle}{\emph{Proceedings of the 2007 ACM conference on Recommender systems}}. \bibinfo{pages}{97--104}.
\newblock


\bibitem[Gao et~al\mbox{.}(2023)]%
        {gao2023echo}
\bibfield{author}{\bibinfo{person}{Yichang Gao}, \bibinfo{person}{Fengming Liu}, {and} \bibinfo{person}{Lei Gao}.} \bibinfo{year}{2023}\natexlab{}.
\newblock \showarticletitle{Echo chamber effects on short video platforms}.
\newblock \bibinfo{journal}{\emph{Scientific Reports}} \bibinfo{volume}{13}, \bibinfo{number}{1} (\bibinfo{year}{2023}), \bibinfo{pages}{6282}.
\newblock


\bibitem[Gillespie et~al\mbox{.}(2013)]%
        {gillespie2014relevance}
\bibfield{author}{\bibinfo{person}{Tarleton Gillespie}, \bibinfo{person}{Pablo~J. Boczkowski}, {and} \bibinfo{person}{Kirsten~A. Foot}.} \bibinfo{year}{2013}\natexlab{}.
\newblock \showarticletitle{The Relevance of Algorithms}.
\newblock In \bibinfo{booktitle}{\emph{Media Technologies: Essays on Communication, Materiality, and Society}}, \bibfield{editor}{\bibinfo{person}{Tarleton Gillespie}, \bibinfo{person}{Pablo Boczkowski}, {and} \bibinfo{person}{Kirsten Foot}} (Eds.). \bibinfo{publisher}{MIT Press}, \bibinfo{address}{Cambridge, MA}, \bibinfo{pages}{167--193}.
\newblock


\bibitem[Guy et~al\mbox{.}(2010)]%
        {guy2010social}
\bibfield{author}{\bibinfo{person}{Ido Guy}, \bibinfo{person}{Naama Zwerdling}, \bibinfo{person}{Inbal Ronen}, \bibinfo{person}{David Carmel}, {and} \bibinfo{person}{Erel Uziel}.} \bibinfo{year}{2010}\natexlab{}.
\newblock \showarticletitle{Social media recommendation based on people and tags}. In \bibinfo{booktitle}{\emph{Proceedings of the 33rd international ACM SIGIR conference on Research and development in information retrieval}}. \bibinfo{pages}{194--201}.
\newblock


\bibitem[Hardt et~al\mbox{.}(2016)]%
        {hardt2016strategic}
\bibfield{author}{\bibinfo{person}{Moritz Hardt}, \bibinfo{person}{Nimrod Megiddo}, \bibinfo{person}{Christos Papadimitriou}, {and} \bibinfo{person}{Mary Wootters}.} \bibinfo{year}{2016}\natexlab{}.
\newblock \showarticletitle{Strategic classification}. In \bibinfo{booktitle}{\emph{Proceedings of the 2016 ACM conference on innovations in theoretical computer science}}. \bibinfo{pages}{111--122}.
\newblock


\bibitem[Haupt et~al\mbox{.}(2023)]%
        {haupt2023recommending}
\bibfield{author}{\bibinfo{person}{Andreas Haupt}, \bibinfo{person}{Dylan Hadfield-Menell}, {and} \bibinfo{person}{Chara Podimata}.} \bibinfo{year}{2023}\natexlab{}.
\newblock \showarticletitle{Recommending to strategic users}.
\newblock \bibinfo{journal}{\emph{arXiv preprint arXiv:2302.06559}} (\bibinfo{year}{2023}).
\newblock


\bibitem[Hu et~al\mbox{.}(2022)]%
        {hu2022interest}
\bibfield{author}{\bibinfo{person}{Xingjian Hu}, \bibinfo{person}{Tiannan Jin}, \bibinfo{person}{Yunjing Lu}, {and} \bibinfo{person}{Shuyu Zhong}.} \bibinfo{year}{2022}\natexlab{}.
\newblock \showarticletitle{The Interest-Based Communities on Xiaohongshu Recreate the Era of “Tribalization”}. In \bibinfo{booktitle}{\emph{2022 6th International Seminar on Education, Management and Social Sciences (ISEMSS 2022)}}. Atlantis Press, \bibinfo{pages}{711--720}.
\newblock


\bibitem[Hu et~al\mbox{.}(2008)]%
        {hu2008collaborative}
\bibfield{author}{\bibinfo{person}{Yifan Hu}, \bibinfo{person}{Yehuda Koren}, {and} \bibinfo{person}{Chris Volinsky}.} \bibinfo{year}{2008}\natexlab{}.
\newblock \showarticletitle{Collaborative filtering for implicit feedback datasets}. In \bibinfo{booktitle}{\emph{2008 Eighth IEEE international conference on data mining}}. \bibinfo{pages}{263--272}.
\newblock


\bibitem[Huang et~al\mbox{.}(2021)]%
        {Huang_2021}
\bibfield{author}{\bibinfo{person}{Yanhua Huang}, \bibinfo{person}{Weikun Wang}, \bibinfo{person}{Lei Zhang}, {and} \bibinfo{person}{Ruiwen Xu}.} \bibinfo{year}{2021}\natexlab{}.
\newblock \showarticletitle{Sliding Spectrum Decomposition for Diversified Recommendation}. In \bibinfo{booktitle}{\emph{Proceedings of the 27th ACM SIGKDD Conference on Knowledge Discovery \& Data Mining}} \emph{(\bibinfo{series}{KDD ’21})}. \bibinfo{publisher}{ACM}.
\newblock
\urldef\tempurl%
\url{https://doi.org/10.1145/3447548.3467108}
\showDOI{\tempurl}


\bibitem[Jacques et~al\mbox{.}(2015)]%
        {jacques2015differentiation}
\bibfield{author}{\bibinfo{person}{Jason~T Jacques}, \bibinfo{person}{Mark Perry}, {and} \bibinfo{person}{Per~Ola Kristensson}.} \bibinfo{year}{2015}\natexlab{}.
\newblock \showarticletitle{Differentiation of online text-based advertising and the effect on users’ click behavior}.
\newblock \bibinfo{journal}{\emph{Computers in Human Behavior}}  \bibinfo{volume}{50} (\bibinfo{year}{2015}), \bibinfo{pages}{535--543}.
\newblock


\bibitem[Jannach et~al\mbox{.}(2018)]%
        {jannach2018recommending}
\bibfield{author}{\bibinfo{person}{Dietmar Jannach}, \bibinfo{person}{Lukas Lerche}, {and} \bibinfo{person}{Markus Zanker}.} \bibinfo{year}{2018}\natexlab{}.
\newblock \showarticletitle{Recommending based on implicit feedback}.
\newblock In \bibinfo{booktitle}{\emph{Social information access: systems and technologies}}. \bibinfo{publisher}{Springer}, \bibinfo{pages}{510--569}.
\newblock


\bibitem[Jawaheer et~al\mbox{.}(2014)]%
        {jawaheer2014modeling}
\bibfield{author}{\bibinfo{person}{Gawesh Jawaheer}, \bibinfo{person}{Peter Weller}, {and} \bibinfo{person}{Patty Kostkova}.} \bibinfo{year}{2014}\natexlab{}.
\newblock \showarticletitle{Modeling user preferences in recommender systems: A classification framework for explicit and implicit user feedback}.
\newblock \bibinfo{journal}{\emph{ACM Transactions on Interactive Intelligent Systems (TiiS)}} \bibinfo{volume}{4}, \bibinfo{number}{2} (\bibinfo{year}{2014}), \bibinfo{pages}{1--26}.
\newblock


\bibitem[Jhaver et~al\mbox{.}(2022)]%
        {jhaver2022designing}
\bibfield{author}{\bibinfo{person}{Shagun Jhaver}, \bibinfo{person}{Quan~Ze Chen}, \bibinfo{person}{Detlef Knauss}, {and} \bibinfo{person}{Amy~X Zhang}.} \bibinfo{year}{2022}\natexlab{}.
\newblock \showarticletitle{Designing word filter tools for creator-led comment moderation}. In \bibinfo{booktitle}{\emph{Proceedings of the 2022 CHI conference on human factors in computing systems}}. \bibinfo{pages}{1--21}.
\newblock


\bibitem[Jhaver et~al\mbox{.}(2023)]%
        {jhaver2023personalizing}
\bibfield{author}{\bibinfo{person}{Shagun Jhaver}, \bibinfo{person}{Alice~Qian Zhang}, \bibinfo{person}{Quan~Ze Chen}, \bibinfo{person}{Nikhila Natarajan}, \bibinfo{person}{Ruotong Wang}, {and} \bibinfo{person}{Amy~X Zhang}.} \bibinfo{year}{2023}\natexlab{}.
\newblock \showarticletitle{Personalizing content moderation on social media: User perspectives on moderation choices, interface design, and labor}.
\newblock \bibinfo{journal}{\emph{Proceedings of the ACM on Human-Computer Interaction}} \bibinfo{volume}{7}, \bibinfo{number}{CSCW2} (\bibinfo{year}{2023}), \bibinfo{pages}{1--33}.
\newblock


\bibitem[Ju et~al\mbox{.}(2008)]%
        {ju2008range}
\bibfield{author}{\bibinfo{person}{Wendy Ju}, \bibinfo{person}{Brian~A Lee}, {and} \bibinfo{person}{Scott~R Klemmer}.} \bibinfo{year}{2008}\natexlab{}.
\newblock \showarticletitle{Range: exploring implicit interaction through electronic whiteboard design}. In \bibinfo{booktitle}{\emph{Proceedings of the 2008 ACM conference on Computer supported cooperative work}}. \bibinfo{pages}{17--26}.
\newblock


\bibitem[Kang and Lou(2022)]%
        {kang2022ai}
\bibfield{author}{\bibinfo{person}{Hyunjin Kang} {and} \bibinfo{person}{Chen Lou}.} \bibinfo{year}{2022}\natexlab{}.
\newblock \showarticletitle{AI agency vs. human agency: understanding human--AI interactions on TikTok and their implications for user engagement}.
\newblock \bibinfo{journal}{\emph{Journal of Computer-Mediated Communication}} \bibinfo{volume}{27}, \bibinfo{number}{5} (\bibinfo{year}{2022}), \bibinfo{pages}{zmac014}.
\newblock


\bibitem[Karizat et~al\mbox{.}(2021)]%
        {karizat2021algorithmic}
\bibfield{author}{\bibinfo{person}{Nadia Karizat}, \bibinfo{person}{Dan Delmonaco}, \bibinfo{person}{Motahhare Eslami}, {and} \bibinfo{person}{Nazanin Andalibi}.} \bibinfo{year}{2021}\natexlab{}.
\newblock \showarticletitle{Algorithmic folk theories and identity: How TikTok users co-produce Knowledge of identity and engage in algorithmic resistance}.
\newblock \bibinfo{journal}{\emph{Proceedings of the ACM on Human-Computer Interaction}} \bibinfo{volume}{5}, \bibinfo{number}{CSCW2} (\bibinfo{year}{2021}), \bibinfo{pages}{1--44}.
\newblock


\bibitem[Kasy and Abebe(2021)]%
        {kasy2021fairness}
\bibfield{author}{\bibinfo{person}{Maximilian Kasy} {and} \bibinfo{person}{Rediet Abebe}.} \bibinfo{year}{2021}\natexlab{}.
\newblock \showarticletitle{Fairness, equality, and power in algorithmic decision-making}. In \bibinfo{booktitle}{\emph{Proceedings of the 2021 ACM Conference on Fairness, Accountability, and Transparency}}. \bibinfo{pages}{576--586}.
\newblock


\bibitem[Kelly and Teevan(2003)]%
        {kelly2003implicit}
\bibfield{author}{\bibinfo{person}{Diane Kelly} {and} \bibinfo{person}{Jaime Teevan}.} \bibinfo{year}{2003}\natexlab{}.
\newblock \showarticletitle{Implicit feedback for inferring user preference: a bibliography}. In \bibinfo{booktitle}{\emph{ACM SIGIR Forum}}, Vol.~\bibinfo{volume}{37}. ACM New York, NY, USA, \bibinfo{pages}{18--28}.
\newblock


\bibitem[Kim and Lim(2023)]%
        {kim2023investigating}
\bibfield{author}{\bibinfo{person}{Hankyung Kim} {and} \bibinfo{person}{Youn-Kyung Lim}.} \bibinfo{year}{2023}\natexlab{}.
\newblock \showarticletitle{Investigating How Users Design Everyday Intelligent Systems in Use}. In \bibinfo{booktitle}{\emph{Proceedings of the 2023 ACM Designing Interactive Systems Conference}}. \bibinfo{pages}{702--711}.
\newblock


\bibitem[Kim et~al\mbox{.}(2011)]%
        {kim2011advertiser}
\bibfield{author}{\bibinfo{person}{Sungchul Kim}, \bibinfo{person}{Tao Qin}, \bibinfo{person}{Hwanjo Yu}, {and} \bibinfo{person}{Tie-Yan Liu}.} \bibinfo{year}{2011}\natexlab{}.
\newblock \showarticletitle{Advertiser-centric approach to understand user click behavior in sponsored search}. In \bibinfo{booktitle}{\emph{Proceedings of the 20th ACM international conference on Information and knowledge management}}. \bibinfo{pages}{2121--2124}.
\newblock


\bibitem[Kim et~al\mbox{.}(2016)]%
        {kim2016click}
\bibfield{author}{\bibinfo{person}{Yoojung Kim}, \bibinfo{person}{Mihyun Kang}, \bibinfo{person}{Sejung~Marina Choi}, {and} \bibinfo{person}{Yongjun Sung}.} \bibinfo{year}{2016}\natexlab{}.
\newblock \showarticletitle{To click or not to click? Investigating antecedents of advertisement clicking on Facebook}.
\newblock \bibinfo{journal}{\emph{Social Behavior and Personality: an international journal}} \bibinfo{volume}{44}, \bibinfo{number}{4} (\bibinfo{year}{2016}), \bibinfo{pages}{657--667}.
\newblock


\bibitem[Klaisubun et~al\mbox{.}(2007)]%
        {klaisubun2007behavior}
\bibfield{author}{\bibinfo{person}{Piyanuch Klaisubun}, \bibinfo{person}{Phichit Kajondecha}, {and} \bibinfo{person}{Takashi Ishikawa}.} \bibinfo{year}{2007}\natexlab{}.
\newblock \showarticletitle{Behavior patterns of information discovery in social bookmarking service}. In \bibinfo{booktitle}{\emph{IEEE/WIC/ACM International Conference on Web Intelligence (WI'07)}}. IEEE, \bibinfo{pages}{784--787}.
\newblock


\bibitem[Klug et~al\mbox{.}(2021)]%
        {klug2021trick}
\bibfield{author}{\bibinfo{person}{Daniel Klug}, \bibinfo{person}{Yiluo Qin}, \bibinfo{person}{Morgan Evans}, {and} \bibinfo{person}{Geoff Kaufman}.} \bibinfo{year}{2021}\natexlab{}.
\newblock \showarticletitle{Trick and Please. A Mixed-Method Study On User Assumptions About the TikTok Algorithm}. In \bibinfo{booktitle}{\emph{Proceedings of the 13th ACM Web Science Conference 2021}} (Virtual Event, United Kingdom) \emph{(\bibinfo{series}{WebSci '21})}. \bibinfo{publisher}{Association for Computing Machinery}, \bibinfo{address}{New York, NY, USA}, \bibinfo{pages}{84–92}.
\newblock
\showISBNx{9781450383301}
\urldef\tempurl%
\url{https://doi.org/10.1145/3447535.3462512}
\showDOI{\tempurl}


\bibitem[Klug et~al\mbox{.}(2023)]%
        {klug2023how}
\bibfield{author}{\bibinfo{person}{Daniel Klug}, \bibinfo{person}{Ella Steen}, {and} \bibinfo{person}{Kathryn Yurechko}.} \bibinfo{year}{2023}\natexlab{}.
\newblock \showarticletitle{How Algorithm Awareness Impacts Algospeak Use on TikTok}. In \bibinfo{booktitle}{\emph{Companion Proceedings of the ACM Web Conference 2023}} (Austin, TX, USA) \emph{(\bibinfo{series}{WWW '23 Companion})}. \bibinfo{publisher}{Association for Computing Machinery}, \bibinfo{address}{New York, NY, USA}, \bibinfo{pages}{234–237}.
\newblock
\showISBNx{9781450394192}
\urldef\tempurl%
\url{https://doi.org/10.1145/3543873.3587355}
\showDOI{\tempurl}


\bibitem[Lee et~al\mbox{.}(2022)]%
        {lee2022algorithmic}
\bibfield{author}{\bibinfo{person}{Angela~Y Lee}, \bibinfo{person}{Hannah Mieczkowski}, \bibinfo{person}{Nicole~B Ellison}, {and} \bibinfo{person}{Jeffrey~T Hancock}.} \bibinfo{year}{2022}\natexlab{}.
\newblock \showarticletitle{The algorithmic crystal: Conceptualizing the self through algorithmic personalization on TikTok}.
\newblock \bibinfo{journal}{\emph{Proceedings of the ACM on Human-Computer Interaction}} \bibinfo{volume}{6}, \bibinfo{number}{CSCW2} (\bibinfo{year}{2022}), \bibinfo{pages}{1--22}.
\newblock


\bibitem[Li et~al\mbox{.}(2022)]%
        {li2022exploratory}
\bibfield{author}{\bibinfo{person}{Nian Li}, \bibinfo{person}{Chen Gao}, \bibinfo{person}{Jinghua Piao}, \bibinfo{person}{Xin Huang}, \bibinfo{person}{Aizhen Yue}, \bibinfo{person}{Liang Zhou}, \bibinfo{person}{Qingmin Liao}, {and} \bibinfo{person}{Yong Li}.} \bibinfo{year}{2022}\natexlab{}.
\newblock \showarticletitle{An exploratory study of information cocoon on short-form video platform}. In \bibinfo{booktitle}{\emph{Proceedings of the 31st ACM International Conference on Information \& Knowledge Management}}. \bibinfo{pages}{4178--4182}.
\newblock


\bibitem[Liang et~al\mbox{.}(2023)]%
        {liang2023enabling}
\bibfield{author}{\bibinfo{person}{Yu Liang}, \bibinfo{person}{Aditya Ponnada}, \bibinfo{person}{Paul Lamere}, {and} \bibinfo{person}{Nediyana Daskalova}.} \bibinfo{year}{2023}\natexlab{}.
\newblock \showarticletitle{Enabling goal-focused exploration of podcasts in interactive recommender systems}. In \bibinfo{booktitle}{\emph{Proceedings of the 28th International Conference on Intelligent User Interfaces}}. \bibinfo{pages}{142--155}.
\newblock


\bibitem[Liu et~al\mbox{.}(2024)]%
        {liu2024train}
\bibfield{author}{\bibinfo{person}{Alexander Liu}, \bibinfo{person}{Siqi Wu}, {and} \bibinfo{person}{Paul Resnick}.} \bibinfo{year}{2024}\natexlab{}.
\newblock \showarticletitle{How to Train Your YouTube Recommender to Avoid Unwanted Videos}. In \bibinfo{booktitle}{\emph{Proceedings of the International AAAI Conference on Web and Social Media}}, Vol.~\bibinfo{volume}{18}. \bibinfo{pages}{930--942}.
\newblock


\bibitem[Liu et~al\mbox{.}(2010)]%
        {liu2010personalized}
\bibfield{author}{\bibinfo{person}{Jiahui Liu}, \bibinfo{person}{Peter Dolan}, {and} \bibinfo{person}{Elin~R{\o}nby Pedersen}.} \bibinfo{year}{2010}\natexlab{}.
\newblock \showarticletitle{Personalized news recommendation based on click behavior}. In \bibinfo{booktitle}{\emph{Proceedings of the 15th international conference on Intelligent user interfaces}}. \bibinfo{pages}{31--40}.
\newblock


\bibitem[Lu et~al\mbox{.}(2015)]%
        {lu2015recommender}
\bibfield{author}{\bibinfo{person}{Jie Lu}, \bibinfo{person}{Dianshuang Wu}, \bibinfo{person}{Mingsong Mao}, \bibinfo{person}{Wei Wang}, {and} \bibinfo{person}{Guangquan Zhang}.} \bibinfo{year}{2015}\natexlab{}.
\newblock \showarticletitle{Recommender system application developments: a survey}.
\newblock \bibinfo{journal}{\emph{Decision support systems}}  \bibinfo{volume}{74} (\bibinfo{year}{2015}), \bibinfo{pages}{12--32}.
\newblock


\bibitem[Lyu et~al\mbox{.}(2024)]%
        {yao2024blind}
\bibfield{author}{\bibinfo{person}{Yao Lyu}, \bibinfo{person}{Jie Cai}, \bibinfo{person}{Anisa Callis}, \bibinfo{person}{Kelley Cotter}, {and} \bibinfo{person}{John~M. Carroll}.} \bibinfo{year}{2024}\natexlab{}.
\newblock \showarticletitle{"I Got Flagged for Supposed Bullying, Even Though It Was in Response to Someone Harassing Me About My Disability.": A Study of Blind TikTokers’ Content Moderation Experiences}. In \bibinfo{booktitle}{\emph{Proceedings of the 2024 CHI Conference on Human Factors in Computing Systems (CHI '24)}}. \bibinfo{publisher}{Association for Computing Machinery}, \bibinfo{address}{New York, NY, USA}, Article \bibinfo{articleno}{741}, \bibinfo{numpages}{15}~pages.
\newblock
\urldef\tempurl%
\url{https://doi.org/10.1145/3613904.3642148}
\showDOI{\tempurl}


\bibitem[{Marketing to China}(2024)]%
        {marketingtochina2024douyin}
\bibfield{author}{\bibinfo{person}{{Marketing to China}}.} \bibinfo{year}{2024}\natexlab{}.
\newblock \bibinfo{title}{Douyin Statistics and Trends}.
\newblock
\newblock
\urldef\tempurl%
\url{https://marketingtochina.com/douyin-statistics-and-trends/}
\showURL{%
\tempurl}
\newblock
\shownote{Accessed: September 1, 2024}.


\bibitem[Mayworm et~al\mbox{.}(2024)]%
        {mayworm2024content}
\bibfield{author}{\bibinfo{person}{Samuel Mayworm}, \bibinfo{person}{Michael~Ann DeVito}, \bibinfo{person}{Daniel Delmonaco}, \bibinfo{person}{Hibby Thach}, {and} \bibinfo{person}{Oliver~L. Haimson}.} \bibinfo{year}{2024}\natexlab{}.
\newblock \showarticletitle{Content Moderation Folk Theories and Perceptions of Platform Spirit among Marginalized Social Media Users}.
\newblock \bibinfo{journal}{\emph{Trans. Soc. Comput.}} \bibinfo{volume}{7}, \bibinfo{number}{1–4}, Article \bibinfo{articleno}{1} (\bibinfo{date}{March} \bibinfo{year}{2024}), \bibinfo{numpages}{27}~pages.
\newblock
\urldef\tempurl%
\url{https://doi.org/10.1145/3632741}
\showDOI{\tempurl}


\bibitem[{Meta}(2022)]%
        {meta_new_2022}
\bibfield{author}{\bibinfo{person}{{Meta}}.} \bibinfo{year}{2022}\natexlab{}.
\newblock \bibinfo{title}{The new {AI}-powered feature designed to improve {Feed} for everyone}.
\newblock
\newblock
\urldef\tempurl%
\url{https://ai.meta.com/blog/facebook-feed-improvements-ai-show-more-less/}
\showURL{%
\tempurl}


\bibitem[Miles and Huberman(1994)]%
        {miles1994qualitative}
\bibfield{author}{\bibinfo{person}{Matthew~B Miles} {and} \bibinfo{person}{A~Michael Huberman}.} \bibinfo{year}{1994}\natexlab{}.
\newblock \bibinfo{booktitle}{\emph{Qualitative data analysis: An expanded sourcebook}}.
\newblock \bibinfo{publisher}{sage}.
\newblock


\bibitem[Namey et~al\mbox{.}(2008)]%
        {namey2008data}
\bibfield{author}{\bibinfo{person}{Emily Namey}, \bibinfo{person}{Greg Guest}, \bibinfo{person}{Lucy Thairu}, \bibinfo{person}{Laura Johnson}, {et~al\mbox{.}}} \bibinfo{year}{2008}\natexlab{}.
\newblock \showarticletitle{Data reduction techniques for large qualitative data sets}.
\newblock \bibinfo{journal}{\emph{Handbook for team-based qualitative research}} \bibinfo{volume}{2}, \bibinfo{number}{1} (\bibinfo{year}{2008}), \bibinfo{pages}{137--161}.
\newblock


\bibitem[Nazari et~al\mbox{.}(2022)]%
        {nazari2022choice}
\bibfield{author}{\bibinfo{person}{Zahra Nazari}, \bibinfo{person}{Praveen Chandar}, \bibinfo{person}{Ghazal Fazelnia}, \bibinfo{person}{Catherine~M Edwards}, \bibinfo{person}{Benjamin Carterette}, {and} \bibinfo{person}{Mounia Lalmas}.} \bibinfo{year}{2022}\natexlab{}.
\newblock \showarticletitle{Choice of implicit signal matters: Accounting for user aspirations in podcast recommendations}. In \bibinfo{booktitle}{\emph{Proceedings of the ACM Web Conference 2022}}. \bibinfo{pages}{2433--2441}.
\newblock


\bibitem[Ngo and Kr{\"a}mer(2021)]%
        {ngo2022exploring}
\bibfield{author}{\bibinfo{person}{Thao Ngo} {and} \bibinfo{person}{Nicole Kr{\"a}mer}.} \bibinfo{year}{2021}\natexlab{}.
\newblock \showarticletitle{Exploring folk theories of algorithmic news curation for explainable design}.
\newblock \bibinfo{journal}{\emph{Behaviour \& Information Technology}} \bibinfo{volume}{41}, \bibinfo{number}{15} (\bibinfo{year}{2021}), \bibinfo{pages}{1--14}.
\newblock


\bibitem[Oard and Kim(2001)]%
        {oard2001modeling}
\bibfield{author}{\bibinfo{person}{Douglas~W Oard} {and} \bibinfo{person}{Jinmook Kim}.} \bibinfo{year}{2001}\natexlab{}.
\newblock \showarticletitle{Modeling Information Content Using Observable Behavior}.
\newblock \bibinfo{journal}{\emph{Proceedings of the Annual Meeting of the American Society for Information Science}}  \bibinfo{volume}{38} (\bibinfo{year}{2001}), \bibinfo{pages}{481--488}.
\newblock


\bibitem[Papadamou et~al\mbox{.}(2022)]%
        {papadamou2022just}
\bibfield{author}{\bibinfo{person}{Kostantinos Papadamou}, \bibinfo{person}{Savvas Zannettou}, \bibinfo{person}{Jeremy Blackburn}, \bibinfo{person}{Emiliano De~Cristofaro}, \bibinfo{person}{Gianluca Stringhini}, {and} \bibinfo{person}{Michael Sirivianos}.} \bibinfo{year}{2022}\natexlab{}.
\newblock \showarticletitle{“It is just a flu”: assessing the effect of watch history on YouTube’s pseudoscientific video recommendations}. In \bibinfo{booktitle}{\emph{Proceedings of the international AAAI conference on web and social media}}, Vol.~\bibinfo{volume}{16}. \bibinfo{pages}{723--734}.
\newblock


\bibitem[Pasquale(2011)]%
        {pasquale2011restoring}
\bibfield{author}{\bibinfo{person}{Frank Pasquale}.} \bibinfo{year}{2011}\natexlab{}.
\newblock \showarticletitle{Restoring transparency to automated authority}.
\newblock \bibinfo{journal}{\emph{J. on Telecomm. \& High Tech. L.}}  \bibinfo{volume}{9} (\bibinfo{year}{2011}), \bibinfo{pages}{235}.
\newblock


\bibitem[P{\'e}rez-Qui{\~n}ones and Sibert(1996)]%
        {perez1996collaborative}
\bibfield{author}{\bibinfo{person}{Manuel~A P{\'e}rez-Qui{\~n}ones} {and} \bibinfo{person}{John~L Sibert}.} \bibinfo{year}{1996}\natexlab{}.
\newblock \showarticletitle{A collaborative model of feedback in human-computer interaction}. In \bibinfo{booktitle}{\emph{Proceedings of the SIGCHI conference on Human Factors in Computing Systems}}. \bibinfo{pages}{316--323}.
\newblock


\bibitem[Petre et~al\mbox{.}(2019)]%
        {petre2019gaming}
\bibfield{author}{\bibinfo{person}{Caitlin Petre}, \bibinfo{person}{Brooke~Erin Duffy}, {and} \bibinfo{person}{Emily Hund}.} \bibinfo{year}{2019}\natexlab{}.
\newblock \showarticletitle{“Gaming the System”: Platform Paternalism and the Politics of Algorithmic Visibility}.
\newblock \bibinfo{journal}{\emph{Social Media + Society}} \bibinfo{volume}{5}, \bibinfo{number}{4} (\bibinfo{year}{2019}), \bibinfo{pages}{2056305119879995}.
\newblock
\urldef\tempurl%
\url{https://doi.org/10.1177/2056305119879995}
\showDOI{\tempurl}


\bibitem[Raza and Ding(2022)]%
        {raza2022news}
\bibfield{author}{\bibinfo{person}{Shaina Raza} {and} \bibinfo{person}{Chen Ding}.} \bibinfo{year}{2022}\natexlab{}.
\newblock \showarticletitle{News recommender system: a review of recent progress, challenges, and opportunities}.
\newblock \bibinfo{journal}{\emph{Artificial Intelligence Review}}  \bibinfo{volume}{55} (\bibinfo{year}{2022}), \bibinfo{pages}{749--800}.
\newblock


\bibitem[Ricks and McCrosky(2022)]%
        {ricks2022does}
\bibfield{author}{\bibinfo{person}{Becca Ricks} {and} \bibinfo{person}{Jesse McCrosky}.} \bibinfo{year}{2022}\natexlab{}.
\newblock \showarticletitle{Does This Button Work? Investigating YouTube’s Ineffective User Controls}.
\newblock \bibinfo{journal}{\emph{Mozilla Foundation}} (\bibinfo{year}{2022}).
\newblock
\urldef\tempurl%
\url{https://foundation. mozilla. org/en/research/library/user-controls/report}
\showURL{%
\tempurl}


\bibitem[Rosenblat and Stark(2016)]%
        {rosenblat2016algorithmic}
\bibfield{author}{\bibinfo{person}{Alex Rosenblat} {and} \bibinfo{person}{Luke Stark}.} \bibinfo{year}{2016}\natexlab{}.
\newblock \showarticletitle{Algorithmic labor and information asymmetries: A case study of Uber’s drivers}.
\newblock \bibinfo{journal}{\emph{International journal of communication}}  \bibinfo{volume}{10} (\bibinfo{year}{2016}), \bibinfo{pages}{27}.
\newblock


\bibitem[Schafer et~al\mbox{.}(1999)]%
        {schafer1999recommender}
\bibfield{author}{\bibinfo{person}{J~Ben Schafer}, \bibinfo{person}{Joseph Konstan}, {and} \bibinfo{person}{John Riedl}.} \bibinfo{year}{1999}\natexlab{}.
\newblock \showarticletitle{Recommender systems in e-commerce}. In \bibinfo{booktitle}{\emph{Proceedings of the 1st ACM conference on Electronic commerce}}. \bibinfo{pages}{158--166}.
\newblock


\bibitem[Seaver(2019)]%
        {seaver2019captivating}
\bibfield{author}{\bibinfo{person}{Nick Seaver}.} \bibinfo{year}{2019}\natexlab{}.
\newblock \showarticletitle{Captivating algorithms: Recommender systems as traps}.
\newblock \bibinfo{journal}{\emph{Journal of material culture}} \bibinfo{volume}{24}, \bibinfo{number}{4} (\bibinfo{year}{2019}), \bibinfo{pages}{421--436}.
\newblock


\bibitem[Serim and Jacucci(2019)]%
        {serim2019explicating}
\bibfield{author}{\bibinfo{person}{Bar{\i}{\c{s}} Serim} {and} \bibinfo{person}{Giulio Jacucci}.} \bibinfo{year}{2019}\natexlab{}.
\newblock \showarticletitle{Explicating "Implicit Interaction" An examination of the concept and challenges for research}. In \bibinfo{booktitle}{\emph{Proceedings of the 2019 CHI conference on human factors in computing systems}}. \bibinfo{pages}{1--16}.
\newblock


\bibitem[Setyani et~al\mbox{.}(2019)]%
        {setyani2019exploring}
\bibfield{author}{\bibinfo{person}{Virda Setyani}, \bibinfo{person}{Yu-Qian Zhu}, \bibinfo{person}{Achmad~Nizar Hidayanto}, \bibinfo{person}{Puspa~Indahati Sandhyaduhita}, {and} \bibinfo{person}{Bo Hsiao}.} \bibinfo{year}{2019}\natexlab{}.
\newblock \showarticletitle{Exploring the psychological mechanisms from personalized advertisements to urge to buy impulsively on social media}.
\newblock \bibinfo{journal}{\emph{International Journal of Information Management}}  \bibinfo{volume}{48} (\bibinfo{year}{2019}), \bibinfo{pages}{96--107}.
\newblock


\bibitem[Shen et~al\mbox{.}(2021)]%
        {shen2021everyday}
\bibfield{author}{\bibinfo{person}{Hong Shen}, \bibinfo{person}{Alicia DeVos}, \bibinfo{person}{Motahhare Eslami}, {and} \bibinfo{person}{Kenneth Holstein}.} \bibinfo{year}{2021}\natexlab{}.
\newblock \showarticletitle{Everyday algorithm auditing: Understanding the power of everyday users in surfacing harmful algorithmic behaviors}.
\newblock \bibinfo{journal}{\emph{Proceedings of the ACM on Human-Computer Interaction}} \bibinfo{volume}{5}, \bibinfo{number}{CSCW2} (\bibinfo{year}{2021}), \bibinfo{pages}{1--29}.
\newblock


\bibitem[Shen et~al\mbox{.}(2012)]%
        {shen2012personalized}
\bibfield{author}{\bibinfo{person}{Si Shen}, \bibinfo{person}{Botao Hu}, \bibinfo{person}{Weizhu Chen}, {and} \bibinfo{person}{Qiang Yang}.} \bibinfo{year}{2012}\natexlab{}.
\newblock \showarticletitle{Personalized click model through collaborative filtering}. In \bibinfo{booktitle}{\emph{Proceedings of the fifth ACM international conference on Web search and data mining}}. \bibinfo{pages}{323--332}.
\newblock


\bibitem[Shepherd(2020)]%
        {SHEPHERD2020102572}
\bibfield{author}{\bibinfo{person}{Ryan~P. Shepherd}.} \bibinfo{year}{2020}\natexlab{}.
\newblock \showarticletitle{Gaming Reddit’s Algorithm: r/the\_donald, Amplification, and the Rhetoric of Sorting}.
\newblock \bibinfo{journal}{\emph{Computers and Composition}}  \bibinfo{volume}{56} (\bibinfo{year}{2020}), \bibinfo{pages}{102572}.
\newblock
\showISSN{8755-4615}
\urldef\tempurl%
\url{https://doi.org/10.1016/j.compcom.2020.102572}
\showDOI{\tempurl}


\bibitem[Simpson and Semaan(2021)]%
        {simpson2021}
\bibfield{author}{\bibinfo{person}{Ellen Simpson} {and} \bibinfo{person}{Bryan Semaan}.} \bibinfo{year}{2021}\natexlab{}.
\newblock \showarticletitle{For You, or For"You"? Everyday LGBTQ+ Encounters with TikTok}.
\newblock \bibinfo{journal}{\emph{Proc. ACM Hum.-Comput. Interact.}} \bibinfo{volume}{4}, \bibinfo{number}{CSCW3}, Article \bibinfo{articleno}{252} (\bibinfo{date}{jan} \bibinfo{year}{2021}), \bibinfo{numpages}{34}~pages.
\newblock
\urldef\tempurl%
\url{https://doi.org/10.1145/3432951}
\showDOI{\tempurl}


\bibitem[Spink and Losee(1996)]%
        {Spink/Losee:96}
\bibfield{author}{\bibinfo{person}{A Spink} {and} \bibinfo{person}{Robert~M Losee}.} \bibinfo{year}{1996}\natexlab{}.
\newblock \showarticletitle{Feedback in {Information} {Retrieval}}.
\newblock \bibinfo{journal}{\emph{Annual Review of Information Science and Technology}} \bibinfo{number}{31} (\bibinfo{year}{1996}), \bibinfo{pages}{33--78}.
\newblock


\bibitem[{Statistica}(2024)]%
        {kuaishou2022user}
\bibfield{author}{\bibinfo{person}{{Statistica}}.} \bibinfo{year}{2024}\natexlab{}.
\newblock \bibinfo{title}{Share of users of Kuaishou (Kwai) in China as of November 2022, by age group}.
\newblock
\newblock
\urldef\tempurl%
\url{https://www.statista.com/statistics/1202735/china-kuaishou-user-age-distribution/#:~:text=As%20of%20November%202022%2C%20almost,living%20in%20China's%20rural%20regions}
\showURL{%
\tempurl}
\newblock
\shownote{Accessed: September 1, 2024}.


\bibitem[Taylor and Choi(2022)]%
        {taylor2022initial}
\bibfield{author}{\bibinfo{person}{Samuel~Hardman Taylor} {and} \bibinfo{person}{Mina Choi}.} \bibinfo{year}{2022}\natexlab{}.
\newblock \showarticletitle{An initial conceptualization of algorithm responsiveness: Comparing perceptions of algorithms across social media platforms}.
\newblock \bibinfo{journal}{\emph{Social Media+ Society}} \bibinfo{volume}{8}, \bibinfo{number}{4} (\bibinfo{year}{2022}), \bibinfo{pages}{20563051221144322}.
\newblock


\bibitem[Van~der Nagel(2018)]%
        {van2018networks}
\bibfield{author}{\bibinfo{person}{Emily Van~der Nagel}.} \bibinfo{year}{2018}\natexlab{}.
\newblock \showarticletitle{"Networks that work too well": intervening in algorithmic connections}.
\newblock \bibinfo{journal}{\emph{Media International Australia}} \bibinfo{volume}{168}, \bibinfo{number}{1} (\bibinfo{year}{2018}), \bibinfo{pages}{81--92}.
\newblock


\bibitem[White et~al\mbox{.}(2005)]%
        {white2005study}
\bibfield{author}{\bibinfo{person}{Ryen~W White}, \bibinfo{person}{Ian Ruthven}, {and} \bibinfo{person}{Joemon~M Jose}.} \bibinfo{year}{2005}\natexlab{}.
\newblock \showarticletitle{A study of factors affecting the utility of implicit relevance feedback}. In \bibinfo{booktitle}{\emph{Proceedings of the 28th annual international ACM SIGIR conference on Research and development in information retrieval}}. \bibinfo{pages}{35--42}.
\newblock


\bibitem[Yi et~al\mbox{.}(2014)]%
        {yi2014beyond}
\bibfield{author}{\bibinfo{person}{Xing Yi}, \bibinfo{person}{Liangjie Hong}, \bibinfo{person}{Erheng Zhong}, \bibinfo{person}{Nanthan~Nan Liu}, {and} \bibinfo{person}{Suju Rajan}.} \bibinfo{year}{2014}\natexlab{}.
\newblock \showarticletitle{Beyond clicks: dwell time for personalization}. In \bibinfo{booktitle}{\emph{Proceedings of the 8th ACM Conference on Recommender Systems}} (Foster City, Silicon Valley, California, USA) \emph{(\bibinfo{series}{RecSys '14})}. \bibinfo{publisher}{Association for Computing Machinery}, \bibinfo{address}{New York, NY, USA}, \bibinfo{pages}{113–120}.
\newblock
\showISBNx{9781450326681}
\urldef\tempurl%
\url{https://doi.org/10.1145/2645710.2645724}
\showDOI{\tempurl}


\bibitem[Zanker and Jessenitschnig(2009)]%
        {zanker2009case}
\bibfield{author}{\bibinfo{person}{Markus Zanker} {and} \bibinfo{person}{Markus Jessenitschnig}.} \bibinfo{year}{2009}\natexlab{}.
\newblock \showarticletitle{Case-studies on exploiting explicit customer requirements in recommender systems}.
\newblock \bibinfo{journal}{\emph{User Modeling and User-Adapted Interaction}}  \bibinfo{volume}{19} (\bibinfo{year}{2009}), \bibinfo{pages}{133--166}.
\newblock


\bibitem[Zhang et~al\mbox{.}(2011)]%
        {zhang2011user}
\bibfield{author}{\bibinfo{person}{Yuchen Zhang}, \bibinfo{person}{Weizhu Chen}, \bibinfo{person}{Dong Wang}, {and} \bibinfo{person}{Qiang Yang}.} \bibinfo{year}{2011}\natexlab{}.
\newblock \showarticletitle{User-click modeling for understanding and predicting search-behavior}. In \bibinfo{booktitle}{\emph{Proceedings of the 17th ACM SIGKDD international conference on Knowledge discovery and data mining}}. \bibinfo{pages}{1388--1396}.
\newblock


\end{thebibliography}

\newpage
\clearpage
\onecolumn
\appendix
\section{Main User Interfaces of Kuaishou and Bilibili Shorts}\label{appendix:interface}
\aptLtoX[graphic=no,type=html]{\begin{figure}[h]
    \centering
    \begin{subfigure}[t]{0.3\textwidth}
        \centering
        \includegraphics[width=0.5\textwidth]{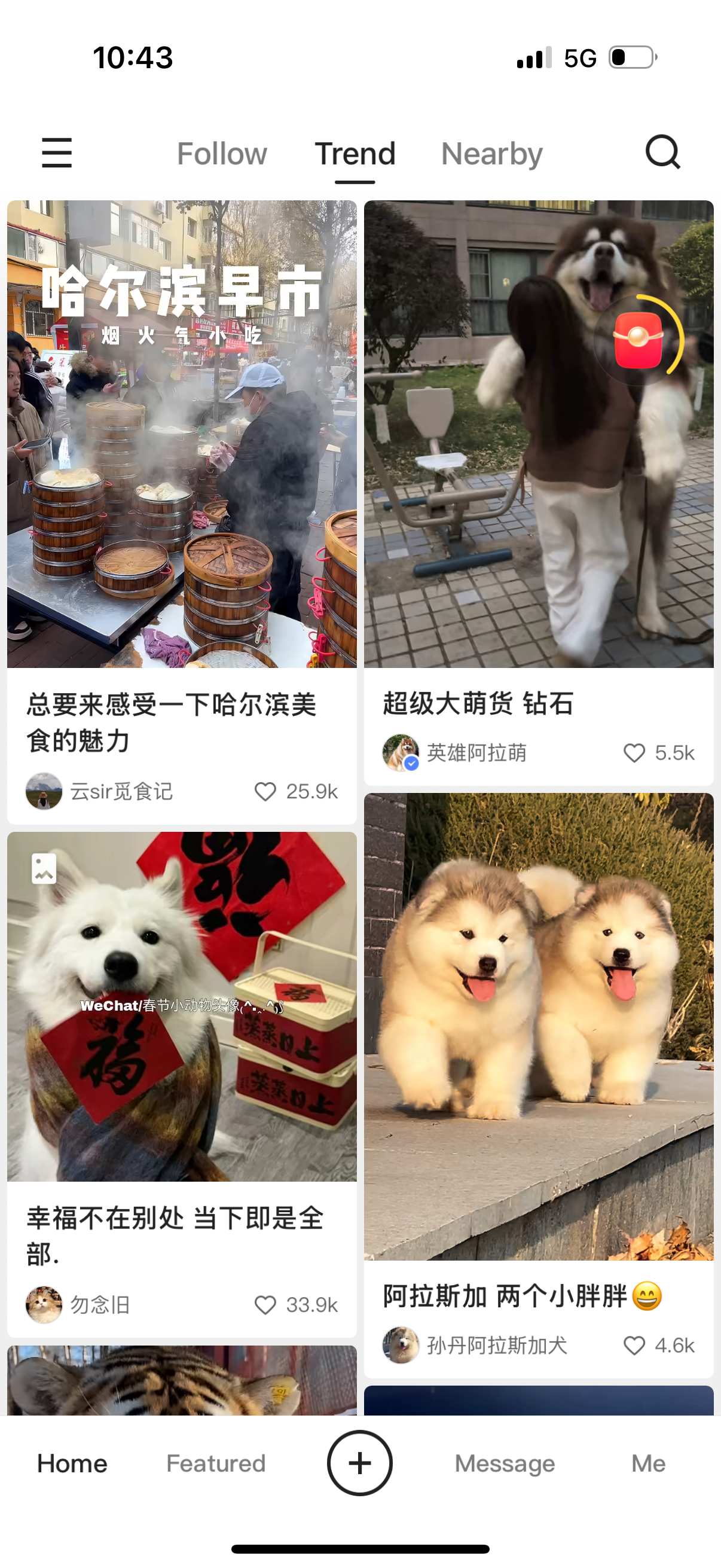}
        \caption{The main page of Kuaishou: ``Trend''}
        \label{fig:ksmain}
    \end{subfigure}
    ~
    \begin{subfigure}[t]{0.3\textwidth}
        \centering
        \includegraphics[width=0.5\textwidth]{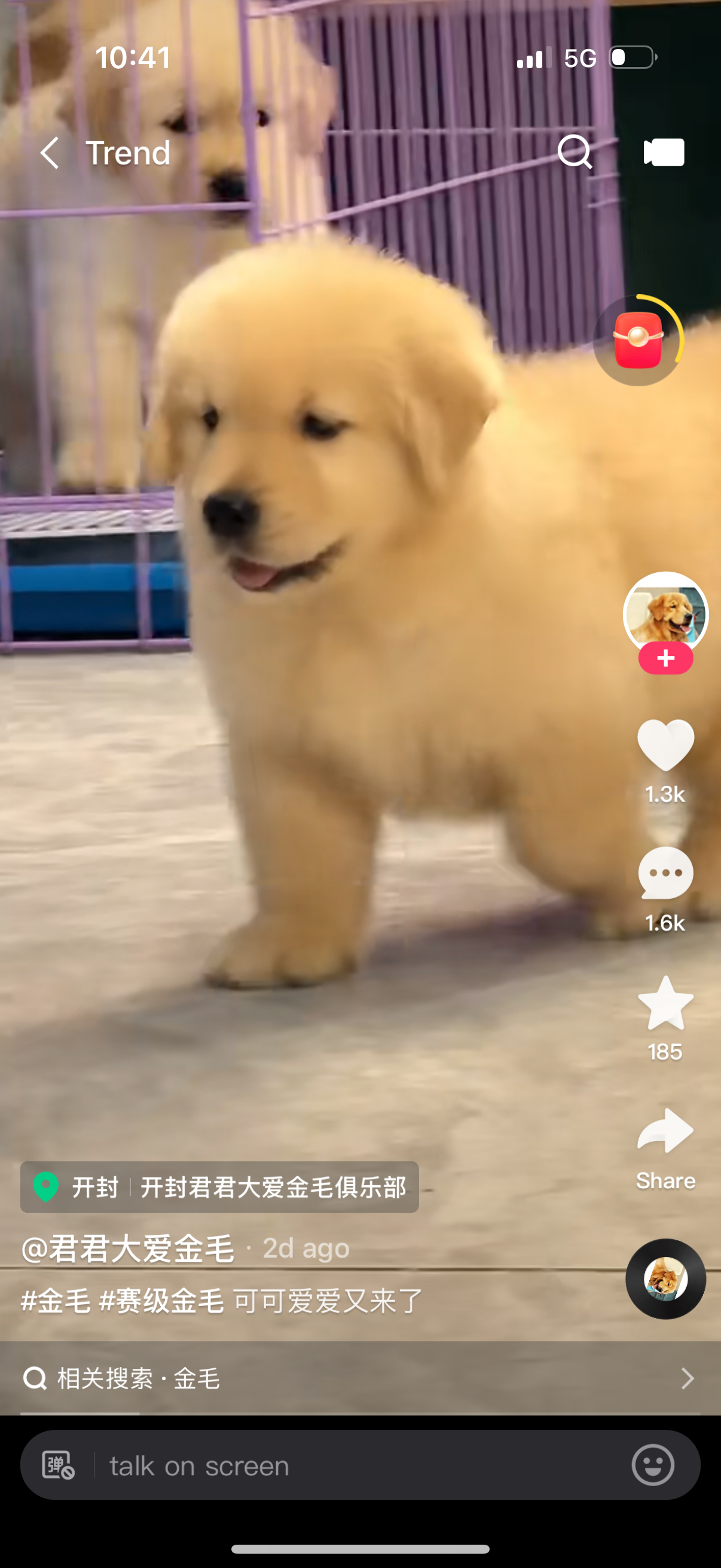}
        \caption{A Kuaishou short video post}
        \label{fig:ksvideo}
    \end{subfigure}
~
    \begin{subfigure}[t]{0.3\textwidth}
        \centering
        \includegraphics[width=0.5\textwidth]{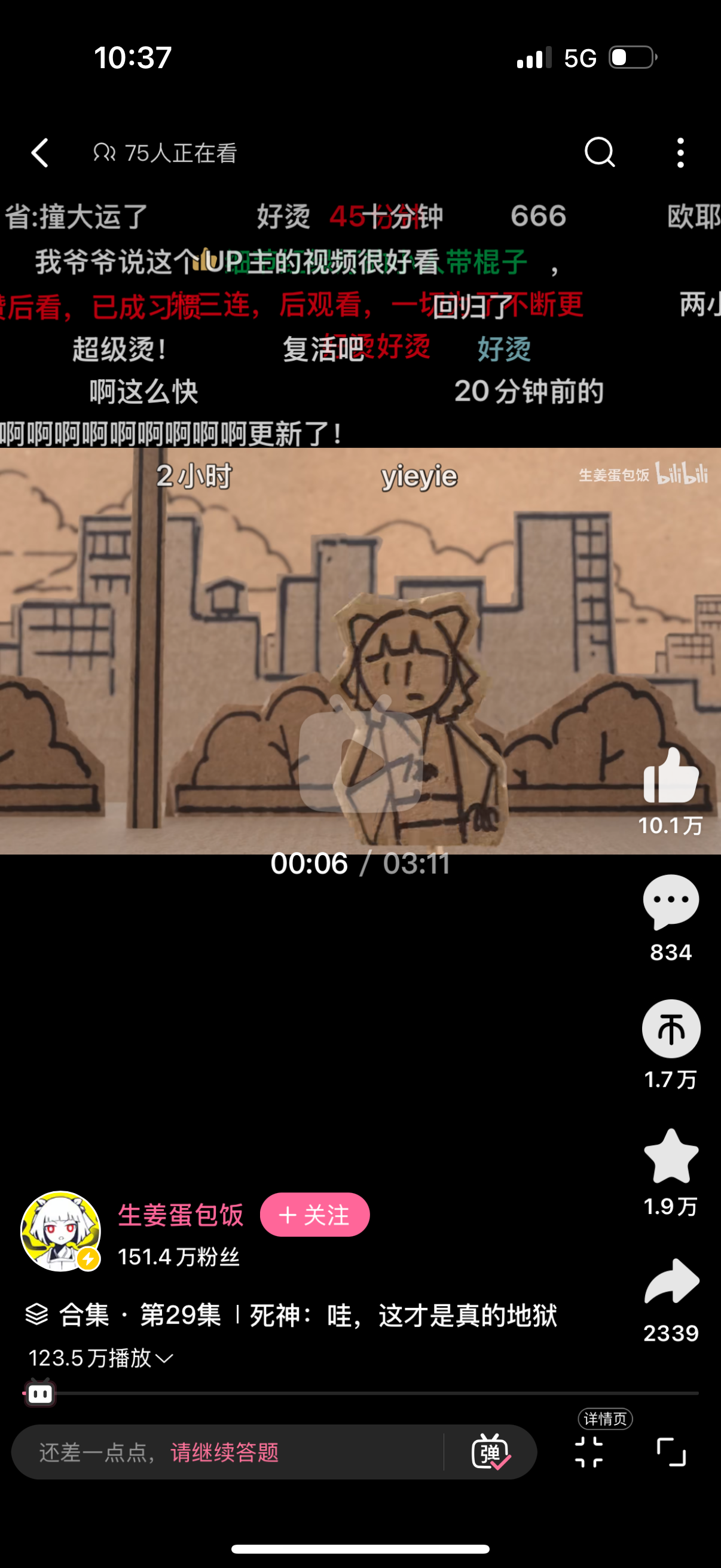}
        \caption{The main interface of Bilibili Shorts}
        \label{fig:bili}
    \end{subfigure}
    \caption{The main user interfaces of Kuaishou and Bilibili Shorts. 
    (a) is the main page of Kuaishou, the ``Trend'' page, which displays a selection of posts recommended by the algorithm. The feed consists mostly of video posts but also includes picture posts, indicated by a picture icon in the top-left corner. This page is similar to the Xiaohongshu ``Explore'' page in Figure~\ref{fig:feedback}.  
    (b) is a short video post page of Kuaishou. Once entering this page, users can scroll vertically to switch video posts, similar to Douyin's ``For You'' page.  
    (c) is the main interface of Bilibili Shorts, similar to the main interface of Douyin.}
    \label{fig:extrainterface}
\end{figure}}{\begin{figure}[h]
    \centering
    \begin{subfigure}[c]{0.3\textwidth}
        \centering
        \includegraphics[width=0.6\textwidth]{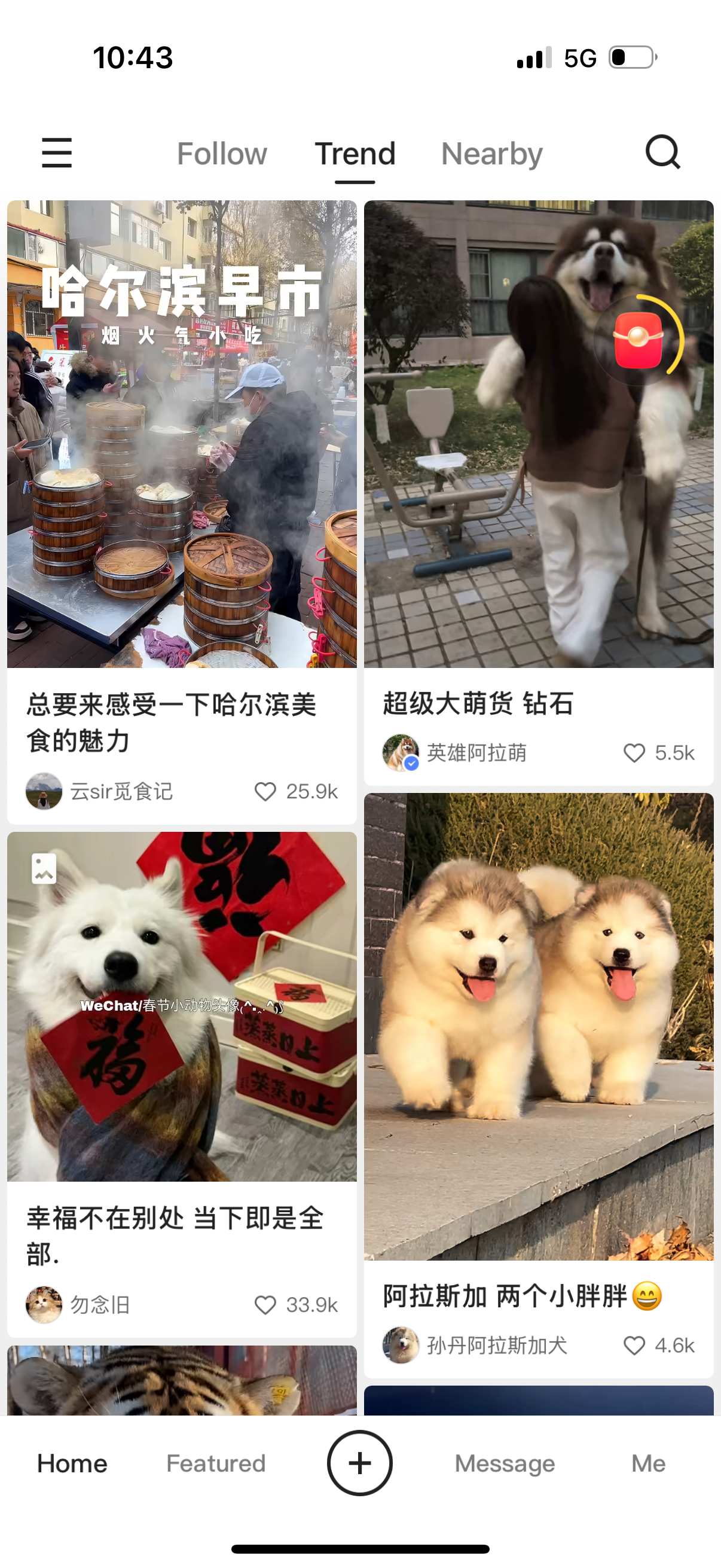}
        \caption{The main page of Kuaishou: ``Trend''}
        \label{fig:ksmain}
    \end{subfigure}
    \hfill
    \begin{subfigure}[c]{0.3\textwidth}
        \centering
        \includegraphics[width=0.6\textwidth]{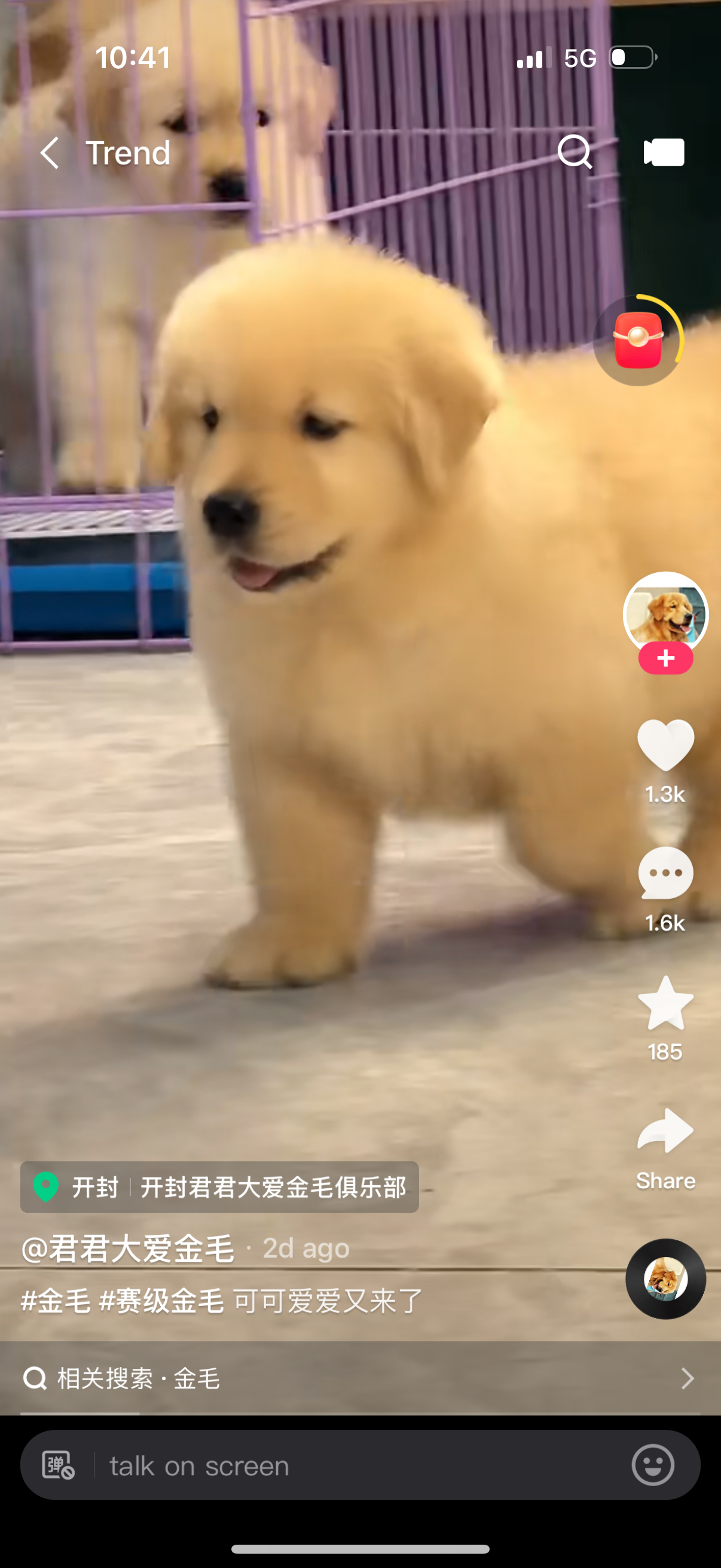}
        \caption{A Kuaishou short video post}
        \label{fig:ksvideo}
    \end{subfigure}
    \hfill
    \begin{subfigure}[c]{0.3\textwidth}
        \centering
        \includegraphics[width=0.6\textwidth]{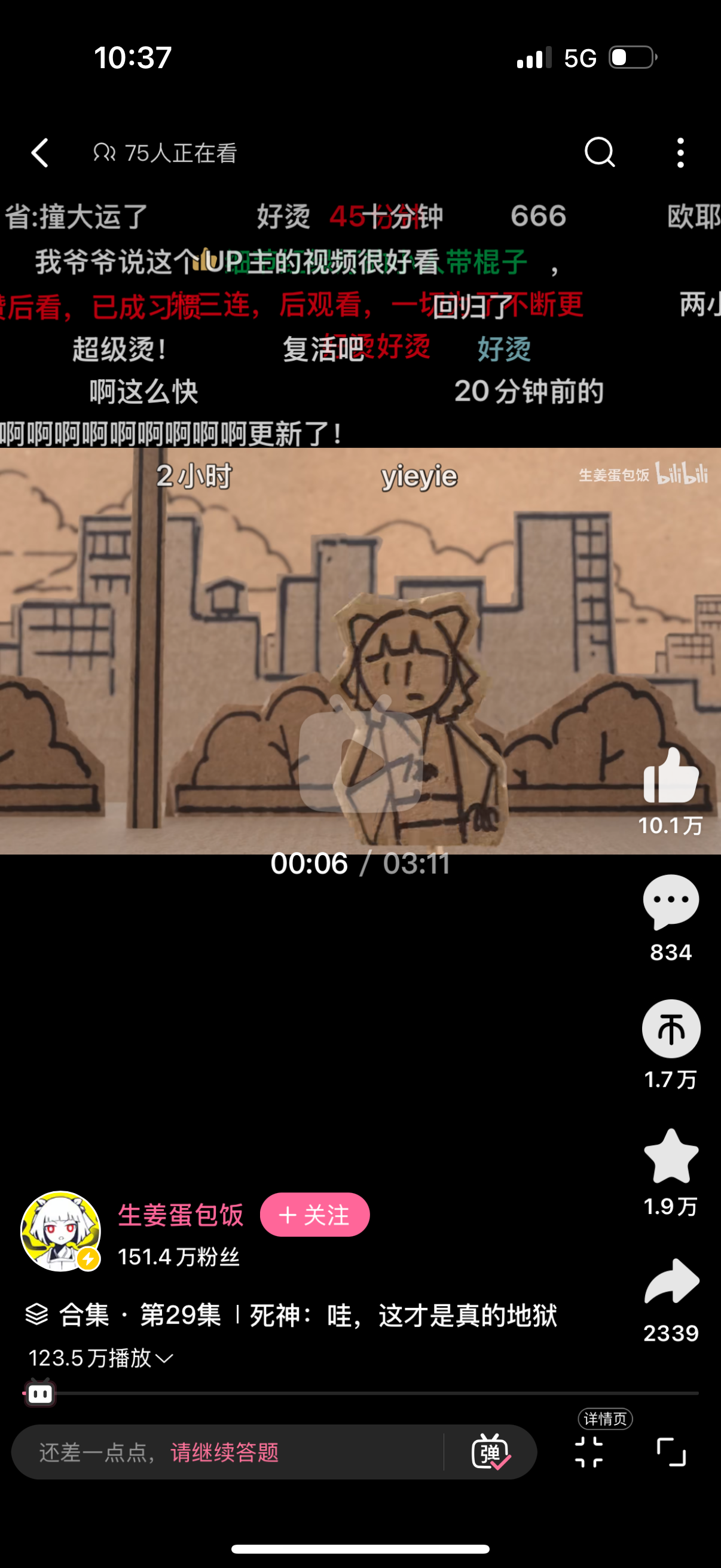}
        \caption{The main interface of Bilibili Shorts}
        \label{fig:bili}
    \end{subfigure}
    \caption{The main user interfaces of Kuaishou and Bilibili Shorts. 
    (\subref{fig:ksmain}) is the main page of Kuaishou, the ``Trend'' page, which displays a selection of posts recommended by the algorithm. The feed consists mostly of video posts but also includes picture posts, indicated by a picture icon in the top-left corner. This page is similar to the Xiaohongshu ``Explore'' page in Figure~\ref{fig:feedback}.  
    (\subref{fig:ksvideo}) is a short video post page of Kuaishou. Once entering this page, users can scroll vertically to switch video posts, similar to Douyin's ``For You'' page.  
    (\subref{fig:bili}) is the main interface of Bilibili Shorts, similar to the main interface of Douyin.}
     \label{fig:extrainterface}
\end{figure}}

\section{Unintentional Implicit Feedback}\label{appendix:unintentional}
\begin{table}[h!]
\small
\caption{Summary of behavior for unintentional implicit feedback and its corresponding characteristics, including platform features involved, the scope of the behavior, the polarity (positive or negative) of the feedback, and the number of participants.}
\label{tab:unintentional}
\resizebox{\textwidth}{!}{%
\begin{tabular}{p{0.37\linewidth}p{0.18\linewidth}p{0.08\linewidth}p{0.14\linewidth}p{0.15\linewidth}}
\toprule
\textbf{Behavior} & \textbf{Feature(s)} &  \textbf{Polarity} & \textbf{Minimum Scope} & \textbf{Participant  Count} \\ \midrule
Share posts with friends& Share to & + & Object & 23 \\
Search for information & Search (cross platform) & + & Class & 21 \\
Collect a post & Collect & + & Object & 21 \\
Create and publish a post & Post & + & Object & 18 \\
Follow a user & Follow & + & Object & 16\\
Expand and view comments & View comments & + & Segment & 15 \\
Comment  & Add comment & + & Segment & 12 \\
View search prompts & Others searched for & + & Segment & 12 \\
Like a post & Like & + & Object & 5\\
Browse a user's profile pages &  & + & Class & 4 \\
Purchase (or from other platforms) & Shop & + & Object & 2 \\
Being at a location & Nearby & + & Class & 2 \\
Save & Save & + & Segment & 1 \\
\bottomrule
\end{tabular}%
}
\end{table}
\clearpage

\section{Participant Demographic Information}\label{appendix:demographics}
\begin{table}[h]
\centering
\caption{Demographic information (i.e., age, gender, education background, and occupation) and platform usage reported by participants.}
\label{tab:participants}
\resizebox{\textwidth}{!}{%
\begin{tabular}{lllllll}
\toprule
No. & Age & Gender & Education & Occupation & Platforms Used & Usage Duration \\\midrule
P01 & 18-25 & Male & Bachelor & Salesperson & Xiaohongshu, Douyin & \textgreater{}4 years \\
P02 & 18-25 & Female & Bachelor & Student & Xiaohongshu, Douyin & \textgreater{}4 years \\
P03 & 26-35 & Female & Bachelor & Administrative staff & Xiaohongshu, Douyin & 3 years \\
P04 & 18-25 & Male & Bachelor & Student & Xiaohongshu, Douyin & 3 years \\
P05 & 18-25 & Female & Bachelor & Student & Xiaohongshu, Douyin, Bilibili & 4 years \\
P06 & 18-25 & Female & Bachelor & Student & Xiaohongshu, Douyin& \textgreater{}4 years \\
P07 & 18-25 & Female & Bachelor & Student & Xiaohongshu, Douyin & 2 years \\
P08 & 18-25 & Female & Bachelor & Student & Xiaohongshu, Douyin & 4 years \\
P09 & 18-25 & Female & Bachelor & Technical professionals & Xiaohongshu, Douyin, Bilibili & \textless{}1 year \\
P10 & 18-25 & Male & Bachelor & Salesperson & Xiaohongshu, Douyin & 3 years \\
P11 & 18-25 & Female & Master & Student & Xiaohongshu,  Douyin, Bilibili & 1 year \\
P12 & 18-25 & Female & Bachelor & Other & Xiaohongshu, Douyin & 3 years \\
P13 & 18-25 & Female & Master & Student & Xiaohongshu, Douyin, Bilibili & \textgreater{}4 years \\
P14 & 18-25 & Female & Bachelor & Student & Xiaohongshu, Douyin & 2 years \\
P15 & 18-25 & Female & Bachelor & Student & Douyin, Xiaohongshu & \textgreater{}4 years \\
P16 & 18-25 & Female & Bachelor & Student & Douyin, Kuaishou & \textgreater{}4 years \\
P17 & 31-40 & Male & Associate & Freelancer & Douyin, Bilibili & \textgreater{}4 years \\
P18 & 18-25 & Female & PhD & Student & Xiaohongshu, Bilibili & 3 years \\
P19 & 18-25 & Female & Master & Student & Kuaishou, Bilibili & 1 year \\
P20 & 26-30 & Male & Associate & Administrative staff & Douyin, Kuaishou & \textgreater{}4 years \\
P21 & 18-25 & Female & Associate & Student & Douyin, Xiaohongshu & \textgreater{}4 years \\
P22 & 31-40 & Male & PhD & Teacher & Douyin, Xiaohongshu & 4 years \\
P23 & 26-30 & Male & PhD & Student & TikTok,  Xiaohongshu & 4 years \\
P24 & 41-50 & Male & Bachelor & Teacher & Douyin, Xiaohongshu, Kuaishou & 4 years \\
P25 & 18-25 & Male & PhD & Student & Xiaohongshu, Bilibili & 4 years \\
P26 & 18-25 & Female & Bachelor & Student & Douyin, Xiaohongshu,  & 4 years \\
P27 & 26-30 & Male & PhD & Student & Douyin, Xiaohongshu & \textgreater{}4 years \\
P28 & 18-25 & Male & Bachelor & Doctor & Douyin, Bilibili & \textgreater{}4 years \\
P29 & 18-25 & Female & Bachelor & Student & Xiaohongshu, Douyin, Kuaishou & 2 years \\
P30 & 31-40 & Male & Bachelor & Non-technical worker & Kuaishou & 1 year \\
P31 & 18-25 & Male & Associate & Non-technical worker & Douyin & \textgreater{}4 years \\
P32 & 51-60 & Male & Technical Secondary School & Non-technical worker & Douyin, Kuaishou & \textgreater{}4 years \\
P33 & 18-25 & Male & Bachelor & Student & Douyin, Bilibili & 2 years \\
P34 & 18-25 & Female & Bachelor & Student & Douyin & \textgreater{}4 years \\\bottomrule
\end{tabular}%
}
\end{table}

\end{document}